\documentclass[pra,twocolumn,preprintnumbers,amsmath,amssymb,superscriptaddress]{revtex4}

\usepackage{color}
\usepackage{graphicx}
\usepackage{epstopdf}
\usepackage{epsfig}
\usepackage{dcolumn}
\usepackage{bm}
\usepackage{hyperref}
\usepackage{mathrsfs}
\usepackage{comment}
\usepackage{soul}
\usepackage[normalem]{ulem} 
\usepackage{ulem}
\usepackage{xcolor}
\usepackage{amsfonts}
\usepackage{amsmath}
\usepackage{amssymb}
\usepackage{mathrsfs}
\usepackage{physics}
\usepackage{hyperref}
\usepackage{soul}
\usepackage{float}
\usepackage{relsize}
\usepackage{placeins}
\usepackage[T1]{fontenc}
\begin{document}


\title{Decoherence effects in quantum nondemolition measurement induced entanglement between Bose-Einstein condensates}

\author{Shuai Gao}
\thanks{These authors contributed equally}
\affiliation{State Key Laboratory of Precision Spectroscopy, School of Physical and Material Sciences,East China Normal University, Shanghai 200062, China} 

\author{Ebubechukwu O. Ilo-Okeke} 
\affiliation{New York University Shanghai, 1555 Century Ave, Pudong, Shanghai 200122, China}
\affiliation{Department of Physics, School of Science, Federal University of Technology, P. M. B. 1526, Owerri, Imo state 460001, Nigeria}

\author{Yuping Mao}
\thanks{These authors contributed equally}
\affiliation{State Key Laboratory of Precision Spectroscopy, School of Physical and Material Sciences,East China Normal University, Shanghai 200062, China}

\author{Manikandan Kondappan}
\affiliation{State Key Laboratory of Precision Spectroscopy, School of Physical and Material Sciences,East China Normal University, Shanghai 200062, China}

\author{Juan E. Aristizabal-Zuluaga}
\affiliation{New York University Shanghai, 1555 Century Ave, Pudong, Shanghai 200122, China}
\affiliation{Grupo de Física Atómica y Molecular, Instituto de Física, Facultad de Ciencias Exactas y Naturales, Universidad de Antioquia UdeA, Calle 70 No. 52-21, Medellín, Colombia}

\author{Valentin Ivannikov}
\affiliation{New York University Shanghai, 1555 Century Ave, Pudong, Shanghai 200122, China} 

\author{Tim Byrnes}
\email{tim.byrnes@nyu.edu}
\affiliation{New York University Shanghai, 1555 Century Ave, Pudong, Shanghai 200122, China} 
\affiliation{State Key Laboratory of Precision Spectroscopy, School of Physical and Material Sciences,East China Normal University, Shanghai 200062, China}
\affiliation{NYU-ECNU Institute of Physics at NYU Shanghai, 3663 Zhongshan Road North, Shanghai 200062, China}
\affiliation{National Institute of Informatics, 2-1-2 Hitotsubashi, Chiyoda-ku, Tokyo 101-8430, Japan}
\affiliation{Department of Physics, New York University, New York, NY 10003, USA}

\date{\today}

\begin{abstract}
We study the robustness of quantum nondemolition (QND) measurement induced entanglement between Bose-Einstein Condensates (BECs). We consider an experimental scheme where two BECs are placed in the paths of a Mach-Zehnder interferometer, and a QND interaction creates entanglement between coherent light and the atoms. We analyze the two dominant channels of decoherence, atomic dephasing and photon loss on the entangled states produced by this scheme. We calculate the effect of dephasing on the variance and expectation values of the spin operators, entanglement and correlation criteria. Our analysis does not use the Holstein-Primakoff approximation, and is capable of modelling long light-atom interaction times, producing non-Gaussian states beyond the two-mode squeezed states. In the presence of dephasing, the entangled states are robust in the macroscopic limit as long as the dimensionless interaction time is less than $ 1/\sqrt{N}$, where $ N $ is the number of atoms in the BEC.  For photon loss, the entangled states generated by long interaction times show a remarkable robustness that makes the scheme promising for various quantum information applications.
\end{abstract}

\maketitle

\section{Introduction}
Quantum mechanics has been traditionally associated with the microscopic world. Since Schr{\"o}dinger's famous gendanken cat experiment, a point that has puzzled physicists is how quantum mechanics at the microscopic level manifests to produce the macroscopic reality described by classical physics \cite{zurek2003decoherence,schlosshauer2007decoherence}. It is now better recognized through an understanding of decoherence that the difficulty in observing quantum phenomena is not only due to the macroscopic nature of the physical system, but the nature of the particular state involved. Certain types of states, such as Schr{\"o}dinger cat states are highly sensitive to decoherence, while other states such as coherent states are relatively insensitive \cite{hornberger2009introduction}.
For an appropriate type of quantum state, it is therefore possible to produce macroscopic quantum states involving a large number of particles.  Examples of these are atomic ensembles \cite{lukin2003colloquium}, Bose-Einstein condensates (BECs) \cite{riedel2010atom,pezze2018quantum}, and micromechanical resonators \cite{o2010quantum}.

Of particular interest are macroscopic states that possess entanglement. The most common type of macroscopic entangled state that is studied are squeezed states, where the uncertainty in one measurement is reduced while increasing the uncertainty of a conjugate variable \cite{sorensen2001many,machida1987observation,shelby1986broad,slusher1986observation,wu1986generation,Schwarzhans1997}. Experimentally this is most widely achieved in optical systems \cite{slusher1985observation,wu1986generation,breitenbach1997measurement} however analogous procedures can be applied to atomic systems such as atomic ensembles which have primarily been considered for applications in quantum metrology \cite{gross2012spin,byrnes2020quantum}. Here, spin squeezed states have been used to improve the sensitivity beyond the standard quantum limit \cite{esteve2008squeezing}. There are a variety of different interactions that exhibit noise reduction in measurement of the internal spin levels such as the interactions of an atomic ensemble inside an optical cavity with either a coherent or optically squeezed light field \cite{Macomber1985a, PhysRevLett.47.709, Wodkiewicz:85, PhysRevA.46.R6797, vernac2000spin, gross2012spin, zhang2014quantum}. Examples of useful squeezing interactions using BECs are the one and two-axis counter-twisting Hamiltonians \cite{kitagawa1993squeezed,hald1999spin,orzel2001squeezed,jo2007long,esteve2008squeezing,bohi2009j,krauter2011entanglement,riedel2010p,riedel2010atom,muessel2014scalable,pezze2018quantum}. In particular quantum nondemolition (QND) measurements have been used as a method of generating squeezing in atomic ensembles \cite{bao2020spin,bao2020retrodiction,hammerer2010quantum,hammerer2004light,appel2009mesoscopic,eckert2008quantum,sewell2012magnetic,koschorreck2010sub,colangelo2017simultaneous,bec1}. Such entangled states have useful applications in quantum metrology 
\cite{giovannetti2011advances,you2017multiparameter,bondurant1984squeezed,gwavesqueeze,Horikiri16,PhysRevA.103.023318}. 

A great majority of the studies regarding entanglement in atomic ensembles and BEC have been for single atomic ensembles. Recently there has been a growing interest in entanglement between two or more atomic ensembles or BECs \cite{bec1,  bec4, bec6}. The first experiments to demonstrate this in atomic ensembles were performed by the Polzik group, using two separated atomic ensembles interacting with a pulse of light performing a non-local Bell measurement on the collective spins. The entanglement was generated between two cesium gas clouds containing $\sim 10^{12}$ atoms each \cite{polzikmacro}. 
Quantum teleportation was successfully demonstrated between two atomic ensembles using discrete \cite{bao2012quantum} and continuous variable encodings \cite{Krauter_2013}. The one- and two-axis spin squeezing states were generalized to the two ensembles case \cite{kurkjian2013spin,li2009spin,byrnes2013fractality,bec4}.  As of yet, entanglement between two spatially separated BECs has not been demonstrated experimentally. However, entanglement has been shown between two spatial regions of the same BEC by generating squeezing and then performing local measurements on parts of the BEC \cite{lange2018entanglement, kunkel2018spatially, fadel2018spatial}. Many BEC entangling schemes have been theoretically proposed using a variety of approaches such as cavity QED \cite{pyrkov2013,hussain2014,abdelrahman2014coherent}, state-dependent forces \cite{treutlein2006}, Rydberg excitations \cite{idlas2016}, and splitting a single squeezed BEC \cite{oudot2017optimal,bec6}.  In particular QND measurement based schemes are promising from the point of view that they have already been achieved in atomic ensembles \cite{oblak2005quantum,di2002entanglement,wang2016schrodinger,julsgaard2001experimental,kuzmich2000generation,chou2005measurement,muschik2011dissipatively,PhysRevLett.85.5643,pettersson2017light,juanQND}. These are fundamental to performing various quantum information tasks based on atomic ensembles, such as quantum teleportation \cite{pyrkov2014quantum}, remote state preparation and clock synchronization \cite{ilo2018remote,manish2021,windpassinger2008nondestructive,meiser2008spin}, and quantum computing \cite{byrnes2012macroscopic}.


In this paper, we calculate the effects of decoherence on entangled BEC states produced by the QND entangling scheme of Ref. \cite{juanQND}. This protocol involves two spatially separated BECs placed in the two arms of a Mach-Zehnder interferometer in the path of coherent light (Fig. 1).  The interaction between light and the BECs occurs via the QND Hamiltonian. The two BECs are then projected onto an entangled state after the measurement of the corresponding optical state.  This is a different geometry to previous schemes (e.g. Ref.  \cite{polzikmacro}) where the atom clouds are placed sequentially, but produce similar effects \cite{pettersson2017light}. 
We investigate the effect of dephasing and photon loss on the entangled BEC state by analyzing quantities such as the variances of spin operators, the probability density distribution and the logarithmic negativity. In particular, we examine states beyond the short time limit where the Holstein-Primakoff approximation is applicable.  Such states are non-Gaussian states with more complex entanglement properties, which are potentially more susceptible to decoherence effects.  Particularly for the case of photon loss, we find that there is a remarkable robustness of the entangled state even for long time scales.    

The structure of this paper is as follows. Sec. \ref{ii} briefly summarizes the QND entangling protocol, defining the physical model and deriving the entangled wavefunction. Sec. \ref{sec:iii} first gives a theory of  dephasing for the entangled state via the Lindblad master equation and shows the numerical results of the dephased state.  Sec. \ref{sec:photonloss} derives the density matrix of the two-pulse state under photon loss, and shows the numerical results of the resulting state. Finally,  The results are then summarized in Sec. \ref{conclusion}. 

\begin{figure}[t]
\includegraphics[width=\linewidth]{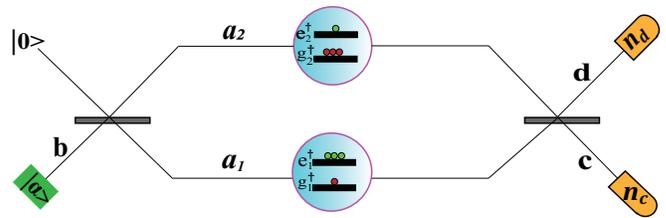}
\setlength{\abovecaptionskip}{-1.5cm}
\caption{\label{juanscheme} Experimental scheme for QND-induced entanglement  between two BECs. A coherent light pulse $|\alpha\rangle $ enters a beam splitter to emerge in two modes $ a_j $, $j \in \{1,2 \}$. The two BECs are placed in the arms of the Mach-Zehnder interferometer. Each optical mode interacts with a BEC through the QND Hamiltonian. Each BEC has two internal states, labeled $e_j, g_j$. The optical modes are then interfered via another beam splitter transforming into two new modes $c$ and $d$, where they are detected with photon numbers $ n_c $ and $ n_d $ respectively. }
\end{figure}

\section{QND Entanglement Protocol} \label{ii}
\subsection{Physical Model}

The QND entangling scheme considered in this paper is shown in Fig. \ref{juanscheme}. Two BECs are each in well-separated traps, for example in separate magnetic traps on an atom chip, or in two optical dipole traps \cite{reichel2011atom,whitlock2009t,abdelrahman2014coherent}. For each BEC, there are two internal energy states which can be populated, with corresponding bosonic annihilation operators $g_j,e_j$, where $j \in \{1,2 \}$ labels each BEC. We can then conveniently define an effective spin using Schwinger boson operators
\begin{align}
\label{schwinger}
S^x_j & = e^{\dag}_j g_j + g^{\dag}_j e_j, \nonumber \\
S^y_j & = -i e^{\dag}_j g_j + i g^{\dag}_j e_j, \nonumber \\
S^z_j & = e^{\dag}_j e_j - g^{\dag}_j g_j, 
\end{align}
which obey the commutation relations $ [S^l, S^m] = 2i \epsilon_{lmn} S^n $, where $ \epsilon_{lmn} $ is the completely anti-symmetric Levi-Civita tensor with $ l,m,n \in \{ x,y,z \} $. We can represent the states of the $ j$th BEC in terms of Fock states
\begin{align}
|k \rangle_{j} = \frac{(e^\dagger_j)^k (g^\dagger_j)^{N-k} }{\sqrt{k! (N-k)!}} | \text{vac} \rangle, 
\label{zfock}
\end{align}
where $ | \text{vac} \rangle $ is the state with no atoms or photons. In Eq. (\ref{zfock}), where appropriate we specify $\ket{k}^{(z)}$ where $(z)$ is the basis as specified in (\ref{schwinger}). The number operator $\hat{N}$ is defined in terms of bosonic creation and annihilation operators
\begin{align}
\hat{N} = g^{\dag}_j g_j + e^{\dag}_j e_j,
\end{align}
returning the occupancy number in the Fock basis
\begin{align}
\hat{N} |k \rangle = N |k \rangle 
\end{align}
as its eigenvalue. The atoms are initially prepared in spin coherent states, which are defined as
\begin{align}
|\theta, \phi\rangle\rangle_j &\equiv \frac{1}{\sqrt{N!}} \left( e^\dag_j \cos \frac{\theta}{2} 
 + e^{i\phi} g^\dag_j \sin \frac{\theta}{2} \right)^N | \text{vac} \rangle \nonumber \\
& = \sum_{k=0}^{N}\sqrt{\binom{N}{k}}\cos^{k}(\frac{\theta}{2})\sin^{N-k}(\frac{\theta}{2})e^{i(N-k)\phi } |k\rangle_j ,
\label{coherent state expression}
\end{align}
where $ 0 \leq \theta \leq \pi$, $ -\pi \leq \phi \leq \pi$ are two arbitrary spherical angles on the Bloch sphere.

The initial state of optical mode $b$ is a coherent state defined as
\begin{equation}
\label{coherent state}
\ket{\alpha}_b \equiv e^{ -\abs{\alpha}^2/2 } e^{\alpha b^\dagger} | \text{vac} \rangle. 
\end{equation}
Mode $b$ then enters a $50\mathbin{:}50$ beam splitter, transforming the mode according to
\begin{align}
b = \frac{1}{\sqrt{2}} ( a_1 + a_2). 
\end{align}
The bosonic annihilation operators for the photonic modes passing through BEC 1 and 2 are $ a_1, a_2 $ respectively. The Stokes operator in the $z$-direction for the optical modes is
\begin{align}
J^z &= a_1^\dag a_1 - a_2^\dag a_2.
\end{align}
The photon number operator is given by
\begin{align}
\hat{n} = a_1^\dag a_1 + a_2^\dag a_2. 
\end{align}
The atoms interact with the light according to the QND Hamiltonian, taking the following form \cite{PhysRevA.73.042112,PhysRevA.79.043815,Echaniz_2005,ilo2014theory,PhysRevA.94.013617}

\begin{align}
H = \frac{\hbar \Omega}{2} \left( S_1^z - S_2^z \right) J^{z}.
\label{Hamiltonian}
\end{align}
Here $\Omega$ is a coupling strength. Note that $J_z$ obeys $\left[ H,J_z\right]=0$ meaning that $J_z$ has no time dependence and the QND Hamiltonian does not affect the $J_z$ observable.  After interacting with the two BECs, a $50\mathbin{:}50$ beam splitter transforms the photon modes according to 
\begin{align}
a_1 & = \frac{1}{\sqrt{2}} ( c + d) \nonumber \\
a_2 & = \frac{1}{\sqrt{2}} ( c - d). 
\label{beamsplitterdef}
\end{align}
Finally, the numbers of photons in modes $c, d$ are then measured. After the measurement the state of the photons in mode $ c $ is
\begin{align}
| n \rangle = \frac{(c^\dagger)^n}{\sqrt{n!}} | \text{vac} \rangle. 
\end{align}
and similarly for mode $ d $.

\subsection{QND Entangled Wavefunction}
We now briefly discuss the quantum state of the BECs after following the protocol in Fig. \ref{juanscheme}. Here we only give the final results, for further details we refer the reader to Ref. \cite{juanQND}. Initially, the state of the BECs are polarized in the $S^x$-direction
\begin{align}
|\Psi_0 \rangle = \left. \ket{\frac{\pi}{2},0} \right\rangle_1 \left. \ket{\frac{\pi}{2},0}\right\rangle_2 ,
\label{becinitial}
\end{align}
and the initial state of photons is given by (\ref{coherent state}).
%
%
Following the experimental scheme, applying a beam splitter, interacting the light via the QND Hamiltonian, applying the second beam splitter and measuring in the photon number basis produces the final state. This derivation can be found in Ref. \cite{juanQND}, and the final state is
\begin{multline}
\ket{{\psi}_{n_c n_d}(\tau)} = \frac{1}{\sqrt{\cal N}} \sum_{k_1, k_2 =0}^N \sqrt{{{N}\choose{k_1}}{{N}\choose{k_2}}} \\
\times C_{n_c n_d}^\alpha [(k_1 - k_2) \tau] \ket{k_1,k_2 }, 
\label{eq:FinaleUnnormalized}
\end{multline}
where we define the dimensionless time $\tau \equiv \Omega t$, and $t$ is the light-atom interaction time. The state 
\begin{align}
| k_1, k_2 \rangle = | k_1 \rangle \otimes | k_2\rangle
\end{align}
is the tensor product state of two Fock states as defined in (\ref{zfock}).  The normalization factor is
\begin{align}
{\cal N} = \sum_{k_1, k_2 =0}^N {{N}\choose{k_1}}{{N}\choose{k_2}} | C_{n_c n_d}[(k_1 - k_2) \tau] |^2 ,   
\end{align}
and we defined 
\begin{align}
C_{n_c n_d}^\alpha (\chi) \equiv \frac{\alpha^{n_c+n_d} e^{-\abs{\alpha}^2/2}}{ \sqrt{n_c ! n_d !} } \cos^{n_c} \chi \sin ^{n_d} \chi. 
\label{cfuncdef}
\end{align}

To see how this results in an entangled state, consider the particular outcome $ n_d = 0 $, which can then be approximated according to Ref. \cite{juanQND} as
\begin{align}
& \ket{\psi_{n_c n_d = 0 }^\text{approx} ( \tau ) } \propto \sum _{k_1, k_2 = 0}^N e^{-\frac{1}{N}\left[(k_1-\frac{N}{2})^2 + (k_2-\frac{N}{2})^2\right]} \nonumber \\
& \times e^{-n_c \tau^2\left(k_1 - k_2\right)^2 / 2} \ket{k_1,k_2 },
\label{approximatewave}
\end{align}
valid for $ N \gg 1 $, $ |\tau| \lesssim 1/\sqrt{N} $, $ |\alpha | \gg 1 $. Here the first exponential factor is a symmetric uncorrelated Gaussian in the Fock state distribution $k_1,k_2$ with averages $ N/2$ and variance  $N/2$. The second exponential factor is a symmetric correlated Gaussian with zero-mean and variance $ \frac{1}{n_c\tau^2}$.  This second factor suppresses all terms except for $ k_1 = k_2 $, and is responsible for producing entanglement between the two BECs. 
%
%
%
%

We note that other methods of QND measurement induced entanglement have been analyzed, often under the Holstein–Primakoff (HP) approximation \cite{kuzmich2000generation,duan2000quantum,julsgaard2001experimental,serafin2021nuclear,tsang2012evading}.  The derivation of the state (\ref{eq:FinaleUnnormalized}) produced by the sequence shown in Fig. \ref{juanscheme} does not require the HP approximation meaning that this state is not constrained to the short time regime $(\tau < \frac{1}{\sqrt{N}})$ \cite{juanQND}. Under the HP approximation, the same procedure produces a two-mode squeezed state such that the EPR variables $\text{Var}(S^x_1 - S^x_2), \text{Var}(S^y_1 + S^y_2) $ are suppressed in comparison to shot noise \cite{julsgaard2001experimental,pettersson2017light}.

\subsection{Two-pulse Scheme}
By introducing a second pulse it is possible to further reduce the variance of the EPR variables $\text{Var}(S^x_1 - S^x_2), \text{Var}(S^y_1 + S^y_2) $  \cite{juanQND}. After obtaining the state (\ref{eq:FinaleUnnormalized}) with the first pulse, we perform a basis rotation such that the $(S_1^y,S_2^y)$ anti-correlations become $(S_1^z,S_2^z)$ correlations. That is, starting from the state (\ref{eq:FinaleUnnormalized}), we apply the transformation $e^{iS_{1}^{x}\pi/4}e^{-iS_{2}^{x}\pi/4} $ which transforms $S_{1}^{y}\rightarrow S_{1}^{z}$ and $S_{2}^{y}\rightarrow -S_{2}^{z}$. We then apply a second optical pulse with the same protocol as the first. Labeling the photon measurement numbers of the first and second rounds as $n_c^{(1,2)},n_d^{(1,2)}$, and interaction times by $\tau_{1,2}$, in Ref. \cite{juanQND} the two-pulse wavefunction was obtained
\begin{align}
    \label{twopulse}
&|\psi_{n_{c}^{(1)} n_{d}^{(1)} n_{c}^{(2)} n_{d}^{(2)}}(\tau_1,  \tau_2)\rangle = \frac{1}{\sqrt{{\cal N}}_2 } \sum_{k_1,k_2,k_1',k_2'=0}^{N}\nonumber \\
&\times \sqrt{\binom{N}{k_{1}'}\binom{N}{k_{2}'}}\langle k_1|e^{iS^{x}\pi/4}
|k_{1}'\rangle \langle k_2|e^{-iS^{x}\pi/4}|k_2'\rangle\nonumber \\
&\times C_{n_{c}^{(1)} n_{d}^{(1)}}^\alpha [(k_{1}' -k_{2}' )\tau_{1}]C_{n_{c}^{(2)}n_{d}^{(2)}}^\alpha [(k_1-k_2)\tau_{2}]|k_1, k_2\rangle,
\end{align}
where $ {\cal N}_2 $ is a normalization factor.  In Sec. \ref{sec:photonloss} we will examine the robustness of the two-pulse state in our analysis. To distinguish between the cases where one and two QND measurements are made, we henceforth refer to the protocol that arrives at  (\ref{eq:FinaleUnnormalized}) as the ``one-pulse'' scheme and (\ref{twopulse}) as the ``two-pulse'' scheme.

\section{Dephasing of the QND Entangled State}\label{sec:iii}

In this section, we examine the effects of $S^{z}$-dephasing decoherence on the entangled state (\ref{eq:FinaleUnnormalized}). Due to the macroscopic nature of our state, dephasing can potentially impact the level of entanglement that can be generated in atomic ensembles. The physical sources that give rise to dephasing in our system are ac Stark scattering and technical noise such as from atom trap current fluctuations \cite{acstark,autler1955stark,reichel2011atom,atomtrapnoise}. In addition, the QND Hamiltonian induces an effective dephasing proportional to the spontaneous decay rate, given by
\begin{align}
    \Gamma = \frac{\Gamma_S|\Omega|^2}{\Delta^2},
    \label{dephasingrateformula}
\end{align}
where $\Gamma$ is the dephasing rate, $\Gamma_S$ is the spontaneous decay rate, $\Omega$ is the laser transition frequency and $\Delta$ is the frequency detuning \cite{acstark}. Current fluctuations in the atom trap randomly shift the energy levels of the logical states, although this effect is suppressed to first order in Zeeman energy if clock states are used.

\subsection{Dephasing Master Equation}
\label{szmaster}

We consider Markovian dephasing to occur on each of the spin states throughout the QND interaction,
\begin{align}
\frac{d\rho }{d t }=  -\frac{\Gamma_{j}}{2}\sum_{n=1}^2\left [ \left ( S_{n}^{j} \right ) ^2\rho -2S_{n}^{j}\rho S_{n}^{j}+ \rho \left ( S_{n}^{j} \right ) ^2 \right ].
\end{align}
Working in units of $ \tau = \Omega t $, we may equally write this as
\begin{align}
\frac{d\rho }{d \tau }=  -\frac{\Gamma_{j}/\Omega}{2}\sum_{n=1}^2\left [ \left ( S_{n}^{j} \right ) ^2\rho -2S_{n}^{j}\rho S_{n}^{j}+ \rho \left ( S_{n}^{j} \right ) ^2 \right ].
\label{Markovian}
\end{align}
We consider the effects of $j \in \{x,z\}$ separately since this can have different effects depending upon the nature of the state \cite{byrnes2013fractality}.  The master equation (\ref{Markovian}) can be solved exactly, such that the density matrix elements evolve according to
\begin{align}
\rho(\tau ) = \langle k_1, k_2 |^{(j)} \rho_0 | k_1' , k_2' \rangle^{(j)} e^{-2 \Gamma \tau/\Omega  [ (k_1 - k_1')^2 + (k_2 - k_2')^2 ]}
\label{rhosol}
\end{align}
for $j \in \{x,z\}$. Here we defined the Fock states in the $ x $-basis as
\begin{align}
| k_1 , k_2 \rangle^{(x)} & = | k_1 \rangle^{(x)} \otimes | k_2 \rangle^{(x)}
\end{align}
where the Fock states in other bases are defined as
\begin{align}
| k \rangle^{(x)} & =  e^{-i S^y \pi/4} | k \rangle^{(z)} 
\label{xbasisfock} \\
| k \rangle^{(y)} & = e^{-i S^z \pi/4} e^{-i S^y \pi/4} | k \rangle^{(z)}
\label{ybasisfock} 
\end{align}
The explicit matrix elements of the transformation are given in Appendix \ref{app:trans}.  

For the initial states, we consider both the QND entangled state  
\begin{align}
    \label{rhoinitial}
    \rho_0 \equiv | \psi_{n_c n_d} (\tau) \rangle \langle  \psi_{n_c n_d}  (\tau) |,
\end{align}
where the state $ | \psi_{n_c n_d}  (\tau) \rangle $ is given in Eq. (\ref{eq:FinaleUnnormalized}).  We also consider similarly the two-pulse initial state (\ref{twopulse}).  Note that in (\ref{rhosol}) we have evolved the Lindblad master equation (\ref{Markovian}) for an equal time $ \tau $ as the entanglement pulse.  We consider this case since for dephasing originating from the ac Stark shift scattering, the dephasing only contributes for the duration of the pulse.  For dephasing originating from technical noise due to the traps, the dephasing times are not necessarily related since the atoms could be held in the traps for longer than the pulse times.  However, in order to minimize decoherence, the state may be measured immediately after the entanglement is generated, and hence this would represent the minimum time that the atoms are in the traps.

%
%
%

Using (\ref{rhoinitial}) as the initial state implies that we perform the dephasing {\it after} the entangled state is prepared. In general, this is an approximation, since dephasing can occur also during state preparation.  For the special case of $S^z$ dephasing, there is no approximation since the Lindblad operation and the QND operation (\ref{Hamiltonian}) commute $[H, S^z] = 0$. On the other hand, since $[H, S^x] \ne 0$, applying $S^x$ dephasing after the interaction to (\ref{rhoinitial}), is only an approximation.  For short QND interaction times the approach becomes more exact, in a similar way that errors in the Trotter approximation are suppressed for short evolution times.  For longer times, 
previous studies with similar entangling operations have demonstrated that the effect on the entanglement gives a similar dependence \cite{byrnes2013fractality, byrnes2012macroscopic, byrnes2015macroscopic}.  To obtain the effect of $S^x$ decoherence we rotate the basis of the initial state according to (\ref{xbasisfock}).

%
%
%
%

\subsection{Expectation values}
\label{means}

\begin{figure}[t]
\includegraphics[width=\columnwidth]{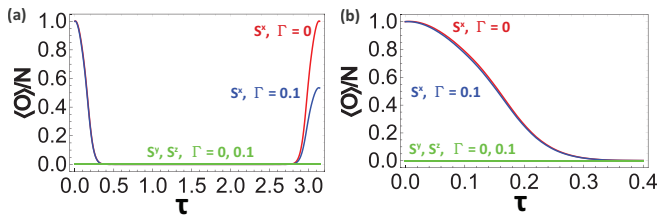}
\caption{\label{Juan expectation with decoherence}Expectation values of the state (\ref{rhosol}) with the state (\ref{eq:FinaleUnnormalized})
and dephasing for the operators $\hat{O} \in \{S^x, S^y, S^z\}$ at long and short time scales. Expectation values are plotted for (a) long time scales; (b) zoomed into short time scales.  Parameters are as marked, and $N=20, {n_c}=100,{n_d}=0$ for all. We set $ \Omega = 1 $ such that the dephasing rate is in units of $ \Omega $, and the time units are $ 1/\Omega $.  }
\end{figure}

We now examine the effect of  $S^z$-dephasing on the expectation values of the state (\ref{rhosol}). The expectation values of an operator $\langle \hat{O} \rangle =\text{Tr}(\hat{O}\rho) $ are evaluated for $\hat{O} \in \{S^x,S^y,S^z\}$. The results are plotted in Fig. \ref{Juan expectation with decoherence}. The expectation values of operators $S^y$ and $S^z$ remain zero regardless of the presence or lack of decoherence. For short time scales, the expectation value of $S^x$ shows very little variation in the presence of decoherence. At time $\tau = \pi$ the expectation value of $S^x$ makes a revival, with the atomic state returning to the initial $ S^x $-polarized state for the $ \Gamma = 0 $ case. The degree of revival is dependent on the value of the decoherence factor $\Gamma$, with a reduced amplitude for $ \Gamma > 0 $.   This shows the direct influence of decoherence at long time scales that is hidden in the HP-approximable $(\tau < \frac{1}{\sqrt{N}})$ regime.

\subsection{Variances}
\label{variances}

\begin{figure}[t]
\centering
\includegraphics[width=\columnwidth]{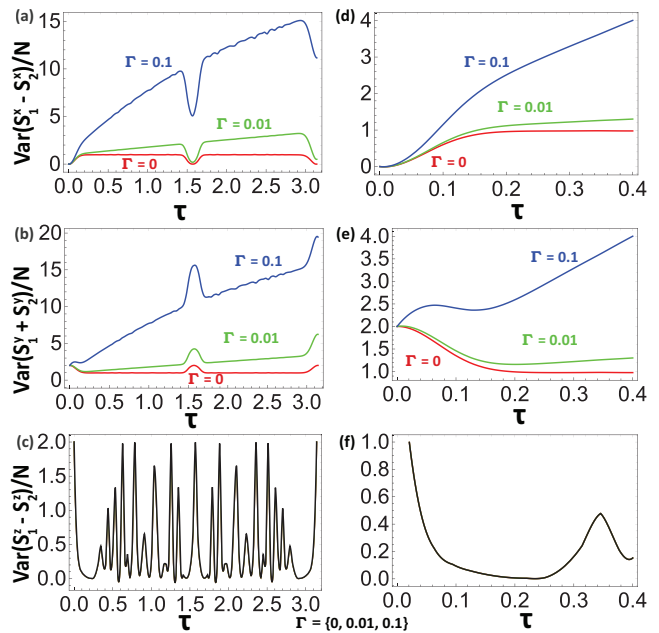}
\caption{The variance of the state (\ref{rhosol}) with the state (\ref{eq:FinaleUnnormalized}) with decoherence for the operators  $S^x_1-S^x_2$, $ S^y_1+S^y_2$ and  $S^z_1-S^z_2$ with $S^z$-dephasing for the rates as marked. Variances are plotted for (a)(b)(c) long time scales; (d)(e)(f) zoomed into short time scales. The parameters $N=20$, ${n_c}=100$, and ${n_d}=0$. Note that for plots (c)(f) all three values of $\Gamma = 0,0.01,0.1$ are shown however there is no difference between each case. We set $ \Omega = 1 $ such that the dephasing rate is in units of $ \Omega $, and the time units are $ 1/\Omega $. }
\label{Juan variance with decoherence}
\end{figure}

Variances of spin-squeezed operators with $S^z$-dephasing are shown as a function of interaction time in Fig. \ref{Juan variance with decoherence} with different decoherence rates $\Gamma$. For zero dephasing $ \Gamma = 0 $ and short times $ 0 < \tau < 1/\sqrt{N} $, the quantities $\text{Var}(S^y_1 + S^y_2)$ and $\text{Var}(S^z_1 - S^z_2)$  show squeezing, with values less than the shot noise level of  $ 2 N $. The variance of $S^y_1 + S^y_2 $ remains lower than shot noise for all times except for $ \tau = n\pi/2 $ where it regains its original value. Meanwhile, the variance of $ S^z_1 - S^z_2 $ undergoes some complex oscillations between low and high values. For $ S^x_1 - S^x_2 $, initially the variance is small as the state is an eigenstate of the observable.  As the interactions are turned on this gradually breaks down as the state departs from being a perfect eigenstate. It was shown in Ref.  \cite{juanQND} that the state possesses correlations between the $ S^x_1,S^x_2 $ hence $\text{Var}(S^x_1 - S^x_2)$ is small for the small time regime. 

With the addition of decoherence, there is an overall increase in the variance for $\text{Var}(S^x_1 - S^x_2)$, $\text{Var}(S^y_1 + S^y_2)$, removing the squeezing effect for sufficiently long times. There is a linear increase in the variance because the dephasing time is set to be equal to the QND interaction time $ \tau $, hence the longer the interaction, the larger the amount of dephasing. The quantity  $\text{Var}(S^z_1 - S^z_2)$ remains at all times unaffected by decoherence since this commutes with the $ S^z $-dephasing.   

%
%

%
%

\begin{figure}[t]
\centering
\includegraphics[width=\columnwidth, trim= {2cm 0 0 1cm}]{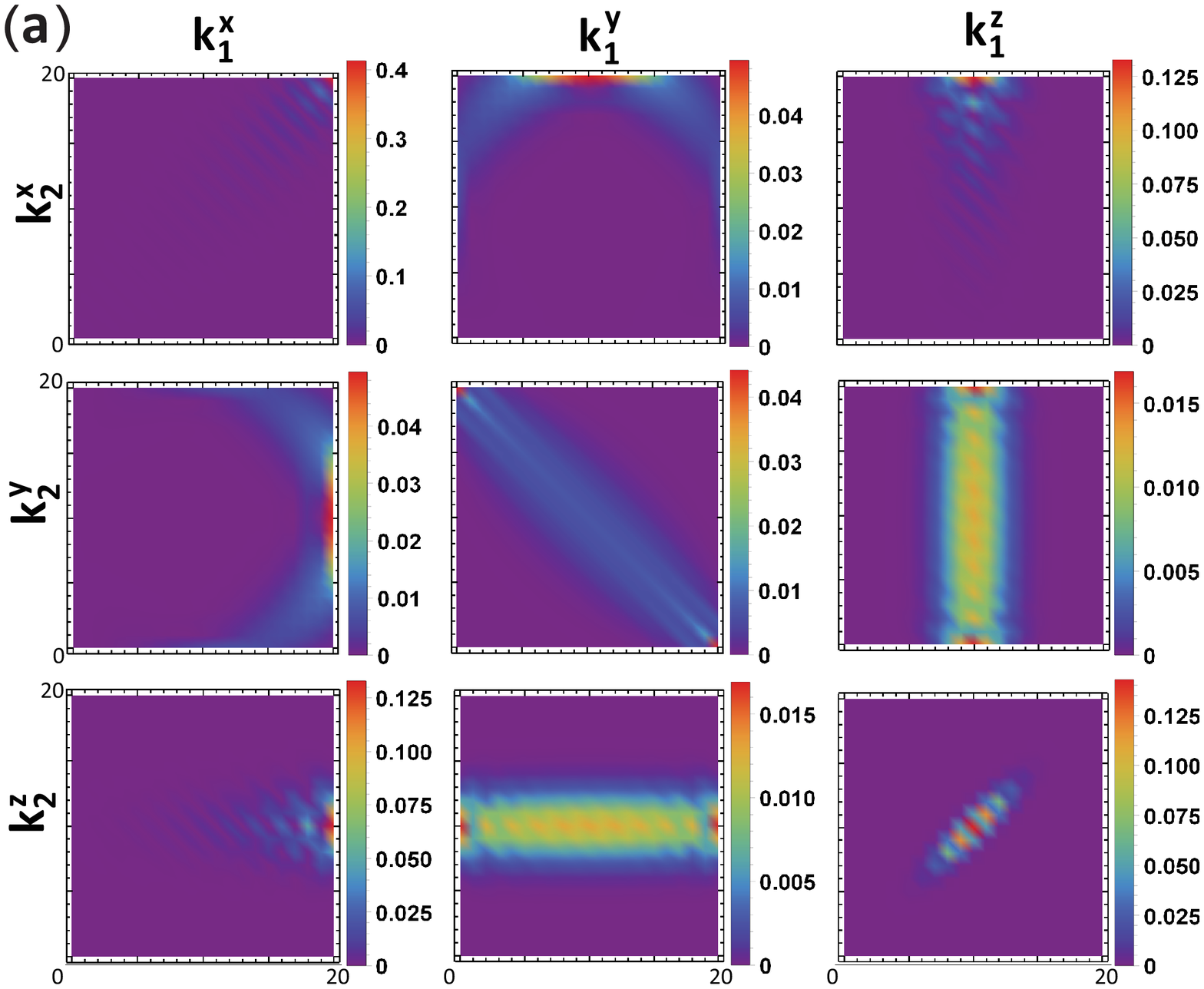}
\includegraphics[width=\columnwidth,trim= {2cm 0 0 0.5cm}]{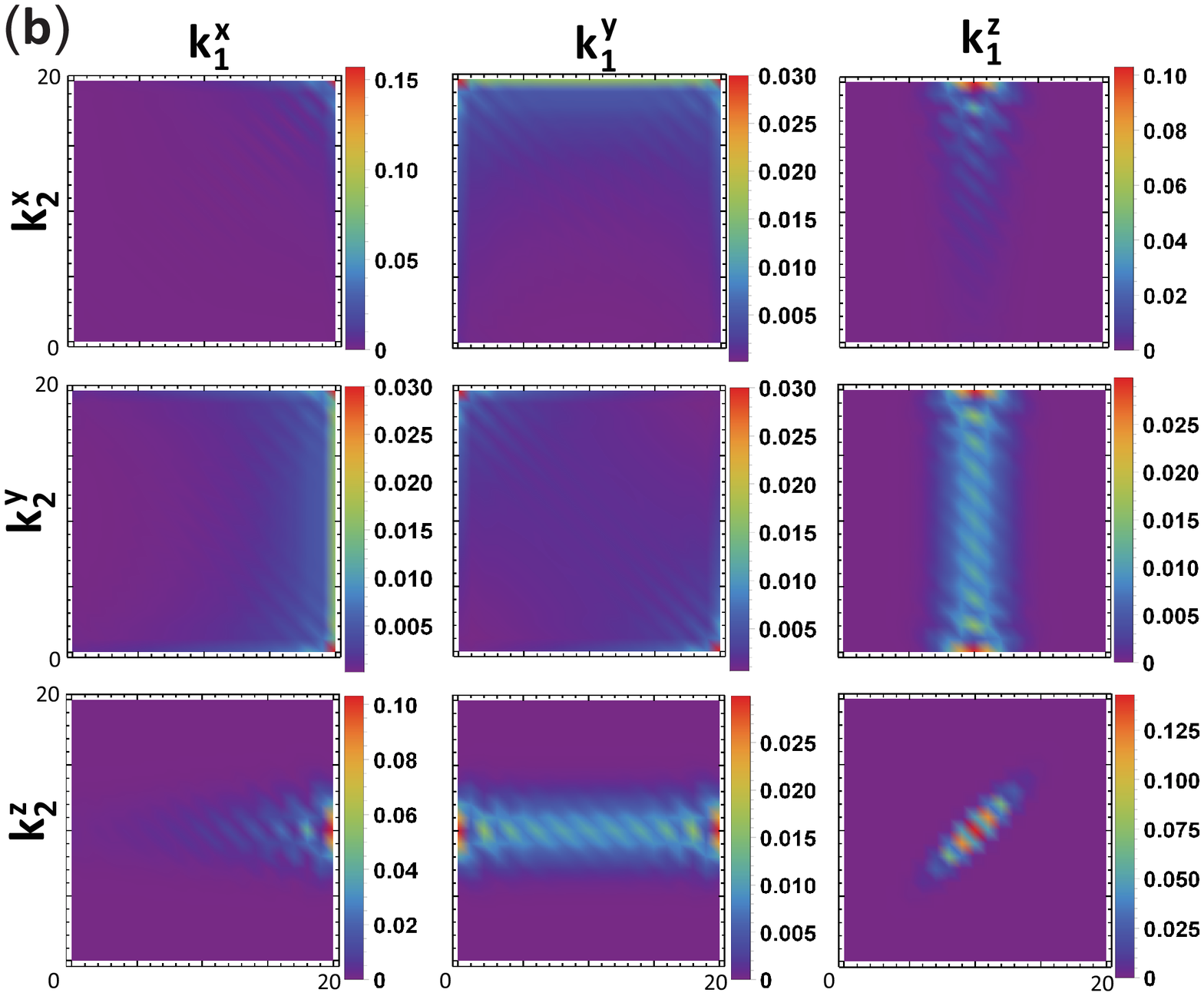}
\caption{\label{Juan_basis_distribution}
Probability distributions after measuring state (\ref{eq:FinaleUnnormalized}) in various bases with parameters $N = 20, n_c = 100, n_d = 0, \hspace{0.1cm} \tau = 0.1 \approx \tau_{\text{opt}}$. Decoherence values for (a)(b) are $\Gamma \in \{0, 1\}$ respectively. Values are for all $\ket{k_j}_{(x)}$ in range $[0, N]$. We set $ \Omega = 1 $ such that the dephasing rate is in units of $ \Omega $, and the time units are $ 1/\Omega $.
}
\end{figure}

\subsection{Probability distribution}
\label{correlation}

We now examine the probability distribution of the state (\ref{rhosol}) in various bases to gain better insight into how the correlations of the state are affected under decoherence. We consider that each BEC is measured in the Fock basis, with probabilities given by 
\begin{align}
p_{i,j}(k_1,k_2)=\langle k_1, k_2 |^{(i,j)} \rho |k_1, k_2\rangle^{(i,j)}. \label{correlation}
\end{align}
where we use the notation
\begin{align}
| k_1, k_2 \rangle^{(i,j)} = | k_1 \rangle^{(i)} \otimes  | k_2 \rangle^{(j)}
\end{align}
and the Fock states in bases $ i, j \in \{ x,y,z \} $ are defined in (\ref{zfock}), (\ref{xbasisfock}), (\ref{ybasisfock}).  
In Fig. \ref{Juan_basis_distribution}(a)(b) we plot all 9 spin combinations of $(i,j)$ with and without the presence of $ S^z $-decoherence. 

First we summarize the decoherence-free case shown in Fig. \ref{Juan_basis_distribution}(a) to understand what the bare correlations are.  For the $ i = j $ cases we see the pattern of correlations ($ i = j = x, z $) and anti-correlations ($ i = j = y $) that was already observed in Fig. \ref{Juan variance with decoherence}.  For the $ i \ne j $ cases, no correlations or anti-correlations are seen in terms of the spin variables. Distributions consist of probabilities centered around $ S^y, S^z = 0 $. 

When decoherence is included, the probability distributions are modified according to Fig. \ref{Juan_basis_distribution}(b).  Since we consider $S^z$-dephasing, the $S_1^z,S_2^z$ components remain unaffected, while the other variables broaden. The general effect is that for the non-commuting variables $S^x, S^y $, the probability distributions are broadened. This results in a reduction of correlation or anti-correlation, increasing the variance as observed in Fig. \ref{Juan variance with decoherence}.

\subsection{Entanglement}
\label{entanglement}

\begin{figure}[t]
\includegraphics[width=\columnwidth]{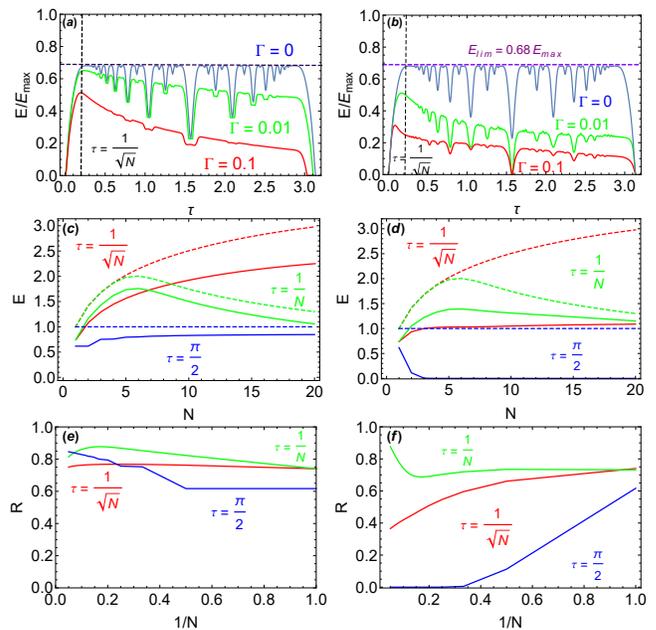}
\caption{\label{Juan Entanglement plot}
Entanglement as quantified by the logarithmic negativity (\ref{lognegdef}) for the state (\ref{rhosol}) in the presence of dephasing.   (a)(b) Entanglement with $S^z$- and $ S^x $-dephasing versus time with decoherence rates as marked and $N=20$, respectively. (c)(d) Entanglement versus $N$ with $S^z$- and $ S^x $-dephasing for the interaction times as marked. Dotted lines show results without dephasing $ \Gamma = 0 $ and solid lines show $ \Gamma = 0.1 $.  (e)(f) Entanglement versus $N$ with $S^x$-dephasing for interaction times as marked. (e)(f) The ratio (\ref{retio of dephasing}) of the dephased entanglement with $ \Gamma = 0.1 $ in comparison with the decoherence-free case $ \Gamma = 0 $ versus $ 1 / N $ for various interaction times as marked.  We set $ \Omega = 1 $ such that the dephasing rate is in units of $ \Omega $, and the time units are $ 1/\Omega $.  }
\end{figure}

Entanglement is a quantity that is highly susceptible to degradation by decoherence, and is one of the central quantities of interest.  In order to quantify the entanglement that is present in the mixed state (\ref{rhosol}), we use logarithmic negativity \cite{vidal2002computable,plenio2005logarithmic} defined as 
\begin{align}
 E={\text {log}}_{2}\left \| \rho^{T_{2}} \right \|, 
 \label{lognegdef}
\end{align}
where $\rho^{T_{2}}$ is the partial transpose on the second BEC. The entanglement ranges from 0 to the maximum value given by
\begin{align}
 E_{\text{max}}=\text {log}_{2}(N+1). 
\end{align}
In Fig. \ref{Juan Entanglement plot}(a) we compare the logarithmic negativity for various values of dephasing rate $ \Gamma$.  As observed in Ref. \cite{juanQND}, the entanglement shows a characteristic fractal devil's crevasse structure. Dips in the entanglement occur at interaction times $ \tau $ that are rational multiples of $ \pi $. This originates from the fact that the QND interaction is of the form $ S^z J^z$, which is the same as the one-axis two-spin squeezing (1A2S) Hamiltonian, which exhibits a similar entanglement dependence \cite{byrnes2013fractality}.  The general effect of the decoherence is to reduce the amount of entanglement with respect to time, with an overall exponential decay profile. For interaction times where the entanglement dips occur, the entanglement appears to be less affected by the decoherence, and do not reduce to the extent that the entanglement ceiling does.  The reason for this is the nature of the state at these points and the fact that we consider $ S^z$-dephasing in Fig. \ref{Juan Entanglement plot}(a).  For example, at $\tau=\pi/2$, the generated state is
\begin{align}
\frac{1}{\sqrt{2}}\left( \lvert \frac{\pi}{2} , 0 \rangle\rangle_1 \lvert \frac{\pi}{2} , 0 \rangle\rangle_2 + \lvert \frac{\pi}{2} , \pi \rangle\rangle_1\lvert \frac{\pi}{2} , \pi  \rangle\rangle_2 \right) ,
\label{catstate}
\end{align}
which is a Bell state consisting of Schrodinger cat states.  Such a state would be expected to be highly sensitive to $ S^x$-dephasing since the Lindblad evolution would strongly affect the sign of the superposition, and quickly make a statistical mixture.  However, $ S^z$-dephasing does not affect the sign of the superposition, and only affects the nature of the spin coherent states involved. Hence we expect that the states at the entanglement dips to be more affected when $ S^x$-dephasing is considered. We see in Fig. \ref{Juan Entanglement plot}(b) that this expectation is confirmed, with the entanglement at $ \tau = \pi/2 $ becoming further reduced by decoherence.  This is generally true of the other entanglement dip times, where the $ S^x $-dephasing degrades the entanglement significantly.

We now examine the entanglement scaling behavior for three characteristic times $\tau \in \{ 1/N ,1/\sqrt{N}, \pi/2 \}$, as shown in Fig. \ref{Juan Entanglement plot}(c)(d) for $ S^z$- and $ S^x$-dephasing respectively.  The time $\tau = 1/N $ corresponds to a time firmly within the HP approximated regime, and exhibits moderate squeezing.  The time $\tau = 1/ \sqrt{N} $ is at the edge of the HP approximable regime and corresponds to timescales near the optimal squeezing time.  Finally, $\tau = \pi/2 $ is deep in the non-Gaussian regime, and the state corresponds to a cat state (\ref{catstate}).  We note similar timescales were examined in Ref. \cite{byrnes2013fractality} to study the robustness of 1A2S states.  
For $\tau=1/N$, within the range of the HP approximation, the entanglement generally survives even as $N$ increases, with a moderate reduction in entanglement.  For $ \tau =   1/\sqrt{N}$, again a reduction in entanglement is seen but it is more strongly affected for the $S^x$-dephasing case. The entanglement is not completely destroyed, however, and appears to reach a finite level. Finally, for $\tau=\pi/2$, the entanglement is rapidly destroyed as $N$ increases in the case that $ S^x$-dephasing is present, but is generally unaffected for $ S^z$-dephasing. This is due to the sensitive nature of the Schr\"odinger cat state that is created at this time.   As was observed in 1A2S states, the time $ \tau =   1/\sqrt{N}$ is a critical time after which the dephasing quite sensitively affects the state.  For the cat-like states at the entanglement dips, dephasing severely affects 
 
To better analyze the robustness of the entanglement as the system size is increased, we examine the ratio 
\begin{align}
R(\Gamma,N)=\frac{E(\Gamma, N)}{E(\Gamma=0, N)}. 
\label{retio of dephasing}
\end{align}
which is a measure of how much the entanglement degrades with respect to the decoherence-free case. In  Fig. \ref{Juan Entanglement plot}(e)(f) we plot the ratio $ R$ for the $ S^z $ and $S^x$-dephasing cases respectively plotted against $ 1/N$ such that the large $ N $ behavior is seen by extrapolating $  1/N \rightarrow 0 $.  For short interaction times $\tau = 1/N $, the ratio appears to approach $ R \approx 1$, meaning that the effect of decoherence becomes increasingly negligible as the system size is increased.  This can be explained by the fact that for large $ N $, the physical interaction time $\tau = 1/N $ becomes shorter, such that the decoherence has less time to act.  At this timescale the states can be considered very robust in the presence of decoherence.  For $\tau = 1/\sqrt{N} $, the ratio $ R $ appears to approach a finite value, or in the worst case go to zero very slowly.  This suggests that entanglement should be observable even for large systems, although there is likely to be some reduction in the entanglement.  Finally, for $\tau = \pi/2 $  the presence of decoherence rapidly destroys any entanglement as the system scales in size.  This occurs for the $ S^x$-dephasing only due to the nature of the particular state that we consider here as explained above.  

In summary, we expect that the states with interaction times at the entanglement dips are most susceptible to decoherence.  The remaining states should be relatively stable in the presence of decoherence.  The reason for the sensitivity to decoherence is due to the fact such dip states are highly symmetric states that are entangled analogs of Schr\"odinger cat states, which are known to be very sensitive to decoherence.  For short times $\tau < 1/\sqrt{N} $, the states are quite robust in the presence of decoherence.

\subsection{Correlation based entanglement criteria}

While logarithmic negativity gives a well-defined quantifiable measure of entanglement for the mixed states that we consider, it requires tomography of the full density matrix, which is often quite demanding in an experimental context due to the large Hilbert space. Another experimental restriction is that only specific types of measurements are typically available. Experimentally, the most convenient observables are low-order expectation values of the total spins. In Ref. \cite{bec4}, it was found that the  Hofmann-Takeuchi criterion  \cite{hofmann2003violation} 
\begin{align}
   {\cal C}_{\text{ent}}    \ge 1  \hspace{1cm} \text{(separable states)}
\label{hofmann}
\end{align}
where
\begin{align}
{\cal C}_{\text{ent}} \equiv & \frac{\text{Var}(S^x_1- S^x_2 ) + \text{Var}(S^y_1+ S^y_2 ) + \text{Var}(S^z_1- S^z_2 ) }{4N}
\end{align}  
gives a convenient and powerful way of detecting entanglement for similarly correlated states.  The Hofmann-Takeuchi criterion is derived for separable states, hence any violation of the inequality signals the presence of entanglement.  On the other hand, a lack of violation gives inconclusive results relating to entanglement. While the type of state that is examined is somewhat different from the bec4 state of Ref. \cite{bec4}, similar correlations are present for small times and we expect that this relation to be a good detector of entanglement.

\begin{figure}[t]
\centering
\includegraphics[width=\columnwidth]{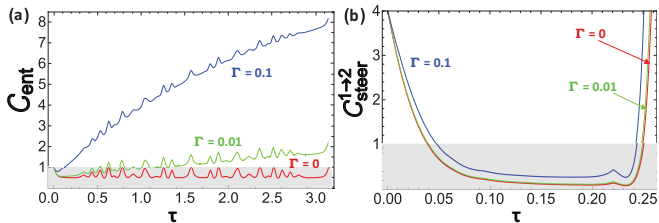}
\caption{\label{correlation entanglement criteria}Detection of entanglement and EPR steering using correlation based criteria. (a) The Hofmann-Takeuchi criterion (\ref{hofmann}) for the state (\ref{rhosol}) with decoherence. (b) The EPR steering detection criterion (\ref{eprsteering}) for the state (\ref{rhosol}) with decoherence. Violations of the inequalities (\ref{hofmann}) and (\ref{eprsteering}) are shown by the shaded area, which corresponds to the detection of entanglement and EPR steering respectively. The chosen parameters are $N=20, n_c=100, n_d=0, \alpha=\sqrt{n_c+n_d}=10$. The decoherence factor choices are as marked.   We set $ \Omega = 1 $ such that the dephasing rate is in units of $ \Omega $, and the time units are $ 1/\Omega $. }
\end{figure}

In Fig. \ref{correlation entanglement criteria}(a) we see the effect of increasing the decoherence rate  $ {\cal C}_\text{ent} $ for a range of interaction times. When $\Gamma=0$, the inequality (\ref{hofmann}) is violated for almost all times, except when $ {\cal C}_\text{ent}  = 1$.  As the decoherence rate increases, only a short initial interaction time region violates the Hoffman-Takeuchi criterion. This reflects the reduction of the correlations, on which the Hoffman-Takeuchi criterion is based on.  We note that this does not necessarily mean that entanglement is not present.  As seen in Fig. \ref{Juan Entanglement plot}(a), in fact entanglement is present for three choices of $ \Gamma $.  The meaning of this is that the criterion (\ref{hofmann}) is simply unable to detect the entanglement, due to the incomplete information of the density matrix.

\subsection{EPR Steering}\label{EPR Steering}
EPR steering describes a phenomenon where one party can non-locally affect the other state through local measurements \cite{Sun2018}. In Ref. \cite{fadel2018spatial,reid2009colloquium}, it was shown  that EPR steering from BEC 1 to BEC 2 exists if it violates the following inequality
\begin{align}
{\cal C}_{\text{steer}}^{1 \rightarrow 2} \equiv \frac{\text{Var}(S^y_1+ S^y_2 )\text{Var}(S^z_1- S^z_2 ) }{\langle S^x_1 \rangle^2} \ge 1. 
\label{eprsteering}
\end{align}
Unlike entanglement or Bell correlations, EPR steering is an intrinsically asymmetric concept. For our state, since our state is symmetric under the interchange of BEC 1 and 2, the quantity is symmetric.  

In Fig. \ref{correlation entanglement criteria}(b) we show the effect of dephasing on the EPR steering criterion (\ref{eprsteering}).  For $\Gamma=0$, the largest violation is seen around $\tau_{\text{opt}}$ coinciding with the time for maximum squeezing. With the introduction of dephasing, we see that the region of violation decreases moderately.  The time regions showing EPR steering are smaller than those showing entanglement detection. This is attributed to EPR steering being narrower criteria than entanglement \cite{adesso2016measures,ma2019operational}.

\section{Photon loss on the QND entangled state}
\label{sec:photonloss}

Another major decoherence channel that can affect the entangled atomic states is photon loss. In the process of interactions between atoms and photons, some of the photons may be lost to the environment, instead of being detected.  Here we analyze the effect of photon loss on the QND entanglement procedure.

\subsection{Density matrix with photon loss}

We again consider that the decoherence acts at the end of the QND entanglement generation process.  Specifically, we assume that the modes $c$ and $d$ have a loss channel before the measurement is performed. The system wavefunction without decoherence at this point is given by Eq. (18) in Ref. \cite{juanQND} which we repeat here for convenience
\begin{align}
|\psi (\tau) \rangle = &  \sum_{k_1, k_2 =0}^N \Psi_{k_1 k_2} \ket{k_1,k_2 } \nonumber \\
& \otimes  | \alpha \cos (k_1 - k_2) \tau \rangle_c  | -i \alpha \sin (k_1 - k_2) \tau \rangle_d .
\label{psibeforemeasurement}
\end{align}
Here $ \Psi_{k_1 k_2} $ is the initial atomic wavefunction, the states of the photon modes are coherent states with modified amplitudes according to the sine and cosine factors.  At this point, the light is still entangled with the atoms.

%
%
%

To consider photon loss it is 
convenient to use the Kraus operator formalism, which is mathematically equivalent to a Lindblad master equation.  Kraus operators for photon loss are defined as
\begin{align}
	A_m=\sum_{n=m}^{\infty}\sqrt{\binom{n}{m}}\sqrt{\gamma ^{n-m}(1-\gamma )^m}|n-m\rangle \langle n|,
\end{align}
where $\gamma$ is the probability of a photon successfully reaching the detector; the probability of photon loss is then $ 1- \gamma$. Applying the Kraus operator on the state (\ref{psibeforemeasurement}) produces 
\begin{align}
& \rho (\tau )  = \sum_{m_1m_2}A_{m_1}\otimes A_{m_2}|\psi (\tau)\rangle\langle \psi (\tau)|A_{m_1}^{\dagger}\otimes A_{m_2}^{\dagger} \nonumber \\
& = \sum_{k_1 k_2 k_{1}' k_{2}'}
\Psi_{k_1 k_2} \Psi_{k_1' k_2'}^* D [ (k_1-k_2 -k_{1}'+k_{2}')\tau ]  \nonumber \\
& \times  |k_1, k_2 \rangle \langle k_1', k_2' |  \otimes | \alpha \cos (k_1 - k_2) \tau \rangle_c   \langle \alpha \cos (k_1 - k_2) \tau  |_c  \nonumber \\ 
&  \otimes 
| -i \alpha \sin (k_1 - k_2) \tau \rangle_d \langle  -i \alpha \sin (k_1 - k_2) \tau |_d .  
\end{align}
Here the $D$-function is a decoherence factor defined as
\begin{align}
D (\chi) = e^{-\left | \alpha \right |^2(1-\gamma) ( 1- \cos \chi )}, 
	\label{Dfunc of photon loss}
\end{align}
which is an exponential originating from the sums in the Kraus operators. Projecting this state on the photonic Fock states gives the state
\begin{align}
& \rho_{n_c n_d} (\tau )	 = \sum_{k_1 k_2 k_{1}' k_{2}'} \Psi_{k_1 k_2} \Psi_{k_1' k_2'}^* D [ (k_1-k_2 -k_{1}'+k_{2}')\tau ]  \nonumber \\
& \times C_{n_cn_d}^{\sqrt{\gamma}\alpha}[(k_1-k_2)\tau]{C^*}_{n_cn_d}^{\sqrt{\gamma}\alpha}[(k_{1}'-k_{2}')\tau] |k_1, k_2 \rangle \langle k_1', k_2' | , 
	\label{density kraus}
\end{align}
where the $ C$-function was defined in (\ref{cfuncdef}). 

We note that in obtaining (\ref{density kraus}) we assumed loss only on modes $ c $ and $ d$. One may ask whether different results would be obtained if the loss occurs on the other modes. In fact, the same state is obtained if the loss is applied on modes $ a_1$ and $ a_2$ instead, as long as the amount of loss on modes $ c $ and $ d $ are equal.  For losses applied on mode $ b $, this simply results in a redefinition of the coherent state amplitude and does not affect the results up to a different value of $ \alpha $.  Finally, after measurement, the photons are disentangled from the atomic wavefunction, and any photon loss is again irrelevant.  In this sense, putting the photon loss on modes $ c $ and $ d$ can be done without loss of generality, for the equal loss case.

\subsection{Variances}

\begin{figure}[t]
	\includegraphics[width=\columnwidth, , trim={35mm 0 0 0}]{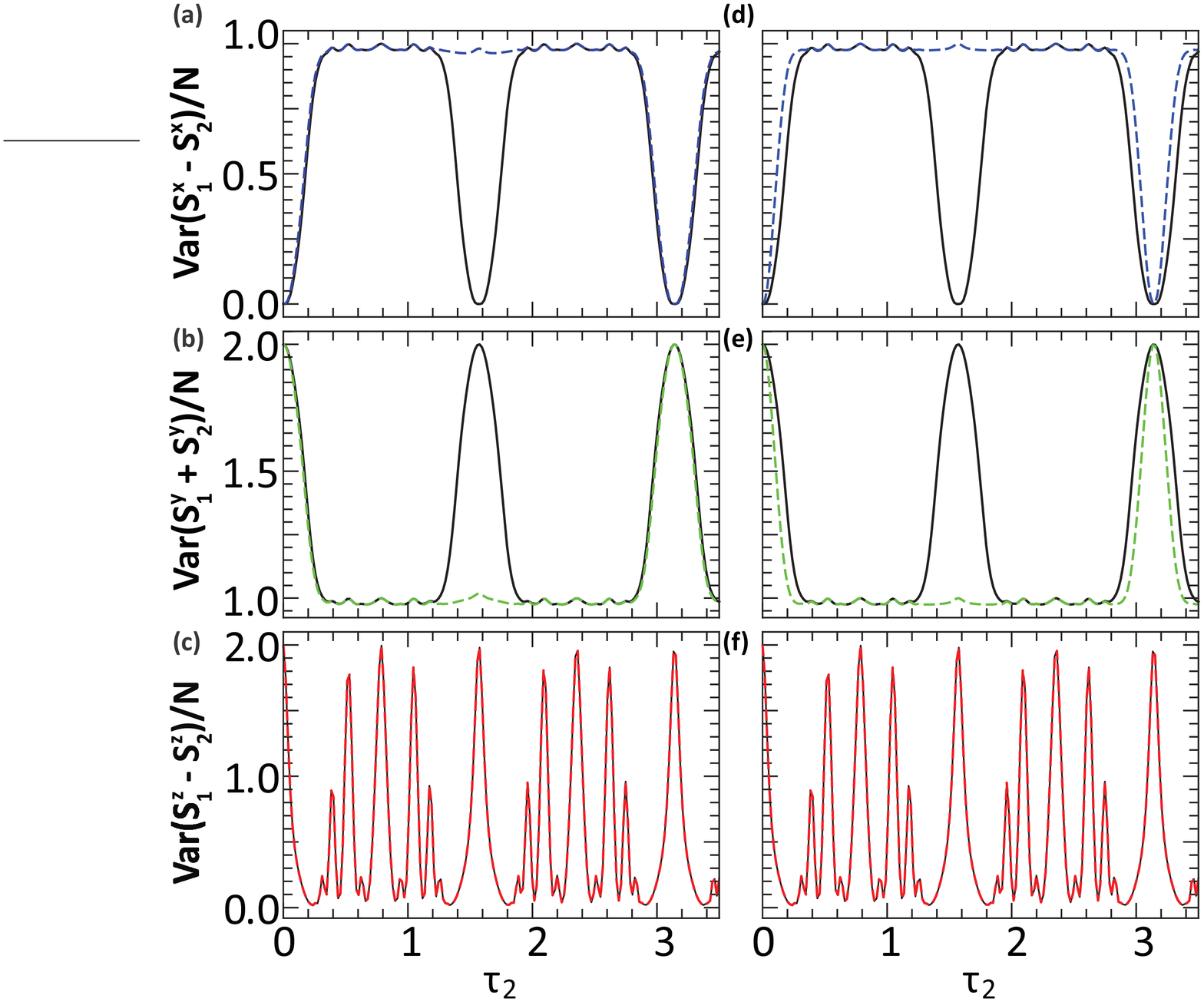}
	\caption{The variance of the entangled state (\ref{density kraus}) for quantities is shown in a two-pulse sequence scheme with photon loss. The solid lines are the variances in the absence of photon loss $\gamma = 1.0$, while the dashed lines are the variances in presence of the photon loss. With $\gamma = 0.9$, and $\gamma = 0.05$ for the first (a)(b)(c) and second (d)(e)(f) columns respectively. The parameters for the figures are $N = 20$, $n^1_c = n^2_c = 20$, $n^1_d = n^2_d = 0$, $\lvert\alpha_1\lvert=\rvert\alpha_2\rvert = \sqrt{20}$, and $\tau_1 = \pi/2$. } 
	\label{fig11}
\end{figure}

Let us now examine particular quantities of interest in the presence of photon loss. In our case, we consider using a rotated version of the state (\ref{eq:FinaleUnnormalized}) as the initial state in (\ref{density kraus}), such that 
\begin{align}
\Psi_{k_1 k_2} = \langle k_1, k_2 | e^{iS^x_1/4} e^{-iS^x_2/4} | \psi_{n_c^{(1)} n_d^{(1)} } (\tau_1) \rangle  .  
\label{initialstatephotonloss}
\end{align}
Note that this state is already a QND entangled state, and this is put into expression (\ref{density kraus}), which performs another QND measurement.  This means that two QND measurements are being performed. The basis rotation is performed such as to further enhance the correlations and anticorrelations in the state, as explained in Ref. \cite{juanQND}.  Without photon loss, the state (\ref{density kraus}) corresponds to the two-pulse scheme wavefunction (\ref{twopulse}).  Since we are considering two QND pulses, there are two interaction times, which we denote as $ \tau_1, \tau_2$ and two sets of measurement outcomes $ n_c^{(1)}, n_d^{(1)}$ and $ n_c^{(2)}, n_d^{(2)}$.  The labels in (\ref{density kraus}) correspond to the second QND measurement.  As such, we henceforth relabel these as $ \tau \rightarrow  \tau_2 $, $ n_c \rightarrow n_c^{(2)}$, and $ n_d \rightarrow n_d^{(2)}$.  

The variances of spin operators for the photon loss state (\ref{density kraus}) are presented in Fig. \ref{fig11}.  
The first QND pulse is applied for a time $ \tau_1 = \pi/2 $ which corresponds to preparing a Bell state consisting of Schrodinger cat states, as given in (\ref{catstate}).  Hence for all values of $ \tau_2$, the state is highly non-Gaussian. We see firstly that $\text{Var}(S^z_1 -S^z_2)$ suffers no effects from photon loss for all times. This is because the $S^z$ operator couples only to diagonal terms for which the decoherence term is 
\begin{align}
D[(k_1 - k_2 - k_1' + k_2')\tau ] \Big|_{k_1'=k_1, k_2'=k_2} = 1.
\end{align}
Hence, the effect of photon loss does not affect $\text{Var}(S^z_1 -S^z_2)$ as shown in Fig. \ref{fig11}(c)(f). 

On the other hand, $\text{Var}(S^x_1 -S^x_2)$ and $\text{Var}(S^y_1 + S^y_2)$ are affected by photon loss. To understand these results, let us first examine these variances without photon loss $\gamma = 1$. When $\tau_2=0$, the state is a well-defined number state in the $S^x$ basis, with no fluctuations and $\text{Var}(S^x_1 -S^x_2) = 0$ as shown by the solid lines in Fig. \ref{fig11}(a)(d). This can be seen explicitly in (\ref{catstate}) since the $ S^x $-polarizations are identical for the two terms in the superposition.  
On the other hand $\mathrm{Var}(S^y_1 - S^y_2)$ and $\mathrm{Var}(S^z_1-S^z_2)$ are both greater than or equal to the shot noise limit for $\tau_2=0$. Applying the second QND measurement $\tau_2>0$, this causes $\text{Var}(S^x_1 -S^x_2)$ to increase from zero to maximum value normalized to unity, while $\text{Var}(S^y_1 + S^y_2)$ decreases to a minimum value normalized to unity. For times $\tau_2=\pi,\,2\pi$, the variances  $\text{Var}(S^x_1 -S^x_2)$ and $\text{Var}(S^y_1 + S^y_2)$ return to their initial value at $\tau_2 = 0$, with $\text{Var}(S^y_1 + S^y_2)$ remaining strictly above unity.
	
In the presence of photon loss $\gamma< 1$, we see almost no effect for  times away from $\tau_2 \approx \pi/2 $, as illustrated by the negligible difference between the dashed and solid lines in Fig. \ref{fig11}. There is little difference between short time variances in \ref{fig11}(a)(b)(c) with $\gamma=0.9$ (i.e. 10\% loss), and in \ref{fig11}(d)(e)(f) $\gamma=0.05$ (i.e. 95\% loss).  $\mathrm{Var}(S^x_1 - S^x_2)$ and $\mathrm{Var}(S^y_1+S^y_2) $ tend to their values without photon loss (solid lines) except in the neighbourhood of $\tau_2 = \pi/2$. $\mathrm{Var}(S^x_1 - S^x_2)$ does not return to zero but rather stays at maximum value that is roughly unity. Similarly, $\mathrm{Var}(S^y_1 - S^y_2)$ remains at its minimum value of unity in the neighbourhood of $\tau_2=\pi/2$. Beyond the neighborhood of $\tau_2 = \pi/2$,  $\text{Var}(S^x_1 -S^x_2)$ and $\text{Var}(S^y_1 + S^y_2)$ return to their values in the absence of photon loss. Note that for the variance to be below unity there must exist correlations in the system. Hence the effect of photon loss is to degrade the correlations in $S^x$ while $\text{Var}(S^y_1 + S^y_2)$ decreases without surpassing the standard quantum limit.
	
We may understand why the photon loss does not affect the correlations to a great degree by studying the decoherence function $ D $. Expanding the cosine in (\ref{Dfunc of photon loss}), we observe that a good approximation can be written as 
\begin{align}
D(\chi) \approx \sum_{n=-\infty}^\infty \exp( - \frac{|\alpha|^2 (1-\gamma) (\chi-2n \pi )^2}{2} ) ,  
\end{align}
valid for $ |\alpha|^2 (1-\gamma) \gg 1 $.  This is a sum of Gaussians which suppresses the amplitude unless $ \chi \approx 2n \pi$.  Examining (\ref{density kraus}) we see that the terms in the sum that do not satisfy 
\begin{align}
 (k_{1}-k_{2}-k_{1}'+k_{2}') \tau=2\pi n
 \label{dfuncond}
\end{align}
will be suppressed. Now notice that the $C$-function (\ref{cfuncdef}) is a Gaussian function  \cite{juanQND} centered at
\begin{align}
(k_1-k_2)\tau =\pm  \arcsin\sqrt{\frac{n_d}{n_d+n_c}} .  
\end{align}
For $n_d=0$, it tends to force $k_1=k_2$ and $k_{1}'=k_{2}'$. This means that even without the $ D$-function, the relation (\ref{dfuncond}) is already satisfied due to the $ C$-functions, and the photon loss does not have a significant effect on the density matrix.  Put in other words, the nature of the state that is produced by the QND interaction coincides with states that are not affected by the photon loss.  For this reason, we do not see a strong effect on the state when photon loss is introduced.


\subsection{Probability distribution}

\begin{figure}[t]
	\includegraphics[width=\columnwidth]{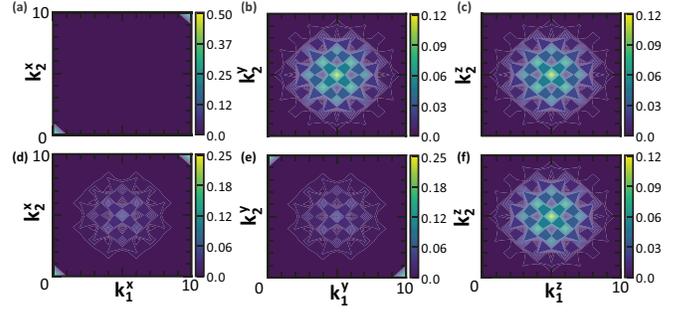}
	\caption{The probability density distribution (\ref{correlation}) for the state (\ref{density kraus}) in the different basis at time $\tau_2=\tau_1 = \pi/2$. Here $ k_{1,2}^{(i)} $ with $ i \in \{x,y,z\} $ indicates the Fock states (\ref{xbasisfock}), (\ref{ybasisfock}), (\ref{zfock}) respectively.  	Parameters used are  (a)(b)(c) $\gamma = 1$  (d)(e)(f) $\gamma = 0.05$. Common parameters are  $N = 10$, $n^1_c = n^2_c = 20$, $n^1_d = n^2_d = 0$, and $\lvert\alpha_1\lvert=\rvert\alpha_2\rvert = \sqrt{20}$} 
		\label{fig13}
\end{figure}

To investigate the further effect of the photon loss on the correlations, we examine probability density on different spin basis $S^x$, $S^y$ and $S^z$. This is defined in the same way as (\ref{correlation}), where the density matrix in this case is given by (\ref{density kraus}).  These probability densities are shown in Fig. \ref{fig13} for the highly non-Gaussian choice of interaction time $ \tau_1 = \tau_2 = \pi/2$ in absence of photon loss ($\gamma = 1$) and 95\% loss ($\gamma =0.05$). For the correlations in the $ S^x $-basis, it is apparent that the effect of photon loss is to cause a loss of correlation in the $S^x$ basis. For the loss-free case we see that a Bell state in the Schrodinger cat basis of the form (\ref{catstate}) is obtained, with $ S^x_1 = S^x_2 = \pm N $.  With the addition of loss, an uncorrelated Gaussian centered at $ S^x_1 = S^x_2 = 0 $ develops.  Since the width of the probability density distribution is related to the variance, the finite width of the distribution accounts for the loss of the correlation that leads to the increase in $\mathrm{Var}(S^x_1 - S^x_2)$ shown in Fig. \ref{fig11}(d). 

For the probability density in the $S^y$ basis as shown in Figs. \ref{fig13}(b)(e), the distribution in the absence of photon loss $\gamma =1$ has large variance as seen by the solid lines of Fig. \ref{fig11}(b)(e). As the photon loss is introduced, a high concentration of the probability density appears for $k_1^y=0, k^y_2=N$ and $k_1^y=N, k^y_2=0$. Thus surprisingly, photon loss in fact produces anti-correlations. The reduction of the variance in Fig. \ref{fig11}(b)(e) with photon loss can be attributed to this distribution.  We interpret this as arising due to photon loss producing an effective $ S^z $-dephasing.  This type of dephasing can be considered an application of a random $S^z$-rotation on both BECs.  The type of correlations as seen in Figs. \ref{fig13}(e) would arise if the state of Figs. \ref{fig13}(a) is rotated by suitable $S^z$-rotations, such that the state (\ref{catstate}) is put in the $S^y$-basis.  

Meanwhile, the probability density in the $S^z$-basis does not show any dependence on the photon loss. This is because the probability density is just the diagonal terms of the density matrix (\ref{density kraus}) for which the $D$ function (\ref{Dfunc of photon loss}) is unity irrespective of the time $\tau$. The independence of this distribution to photon loss supports the interpretation that the photon loss produces an effective $S^z$-dephasing.

\subsection{Entanglement}

\begin{figure}[t]
\includegraphics[width=\columnwidth, trim={5mm 5mm 0mm 5mm}]{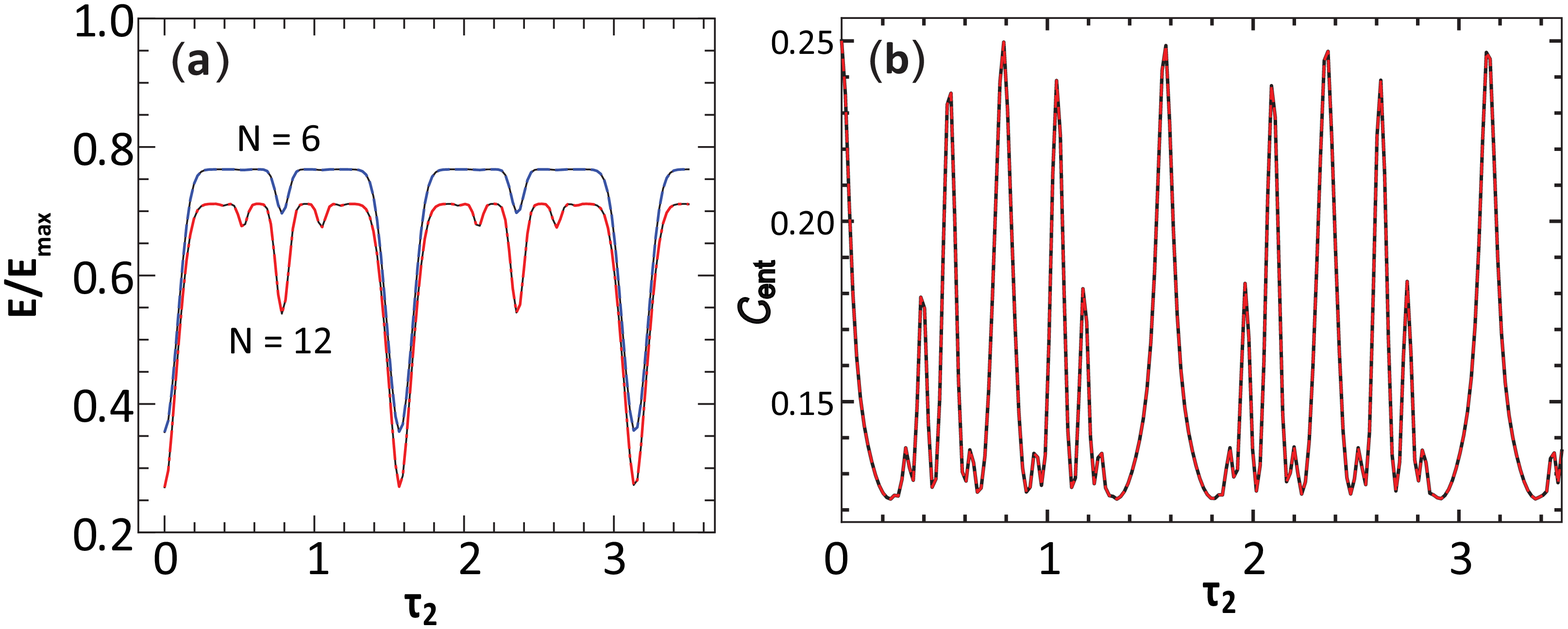}
	\caption{(a) The normalized logarithmic negativity (\ref{lognegdef}) of the two-pulse scheme density matrix (\ref{density kraus}) in the presence of photon loss.  The atom number $N $ is as marked and the  photon detection probabilities $\gamma \in \{1, 0.9\}$ are solid and dashed lines respectively. The parameters are $\lvert\alpha_1\rvert = \lvert\alpha_2\rvert = \sqrt{20}$, $n_c^1=n_c^2 = 20$, $n^1_d = n^2_d=0$, and $\tau_1=\pi/2$. 
	(b) Entanglement detection using the Hofmann-Takeuchi criterion for entangled BECs in the presence of photon loss. The solid lines are the criterion in the absence of photon loss $\gamma = 1$, while the dashed lines are the criterion in presence of the photon loss $\gamma = 0.05$. Parameters are $N = 20$, $n^1_c = n^2_c = 20$, $n^1_d = n^2_d = 0$, $\lvert\alpha_1\lvert=\rvert\alpha_2\rvert = \sqrt{20}$, and $\tau_1 = \pi/2$.
	} 
	\label{fig14}
\end{figure}

We now examine the entanglement in the state (\ref{density kraus}) by plotting the logarithmic negativity in Fig. \ref{fig14}(a).  We see that from the perspective of entanglement, photon loss has a negligible impact on the entanglement present in the quantum states of the BECs for all interaction times. From the effect of the photon loss on the variances as seen in Fig. \ref{fig11}, we expect the entanglement to be relatively unaffected at all times. We may understand this by considering that the effect of photon loss can be viewed as a similar effect to $ S^z $-dephasing, since it does not affect the $ S^z $ probability distribution, as seen in Fig. \ref{fig13}(c)(f).  If we apply a random $ S^z $-rotation to the state (\ref{catstate}), then the entanglement is unaffected since this only has the effect of redefining the spin coherent states that are involved. In the original state (\ref{catstate}), all spin coherent states are $ S^x$-eigenstates, but under a $S^z$-rotation these may become $ S^y$-eigenstates as seen in Fig. \ref{fig13}(e).  Hence although the nature of the entangled state changes under photon loss, the entanglement is unaffected.

\subsection{Correlation based entanglement criteria}

We finally examine the effect of photon loss on the   Hoffman-Takeuchi entanglement detection criterion (\ref{hofmann}).  The results are shown in Fig. \ref{fig14}(b). Remarkably, as was observed with the logarithmic negativity, there is no discernible difference in the correlations even with 95\% photon loss. We find that (\ref{hofmann}) is violated and is well below unity. This violation implies that there is entanglement between the two separated BEC states even in the presence of photon loss. It also suggests that the loss of photons does not impact the quality of the correlations as long as a sufficiently large coherent state population survives at detection $ |\alpha|^2\gamma \gg 1$.

\section{Experimental parameters}
Here we give some experimental parameters to illustrate how the QND entangling procedure may be realized. One possible way to implement our scheme is using trapped cold gas of $^{87} \text{Rb} $ atoms, where the bosonic operators $ g_j, e_j $ of (\ref{schwinger}) correspond to the hyperfine ground states $F = 1,m_F =-1$ and $ F = 2,m_F=+1$ clock states respectively \cite{riedel2010atom,pezze2018quantum,fadel2018spatial,bohi2009coherent}. The QND Hamiltonian (\ref{Hamiltonian}) is then produced by a second order off-resonant transition from the ground state to an excited state, using a D$_2 $ transition, for example \cite{schleier2010states}.  The QND coupling is given by   
\begin{align}
\Omega = \frac{\Omega^2_R}{\Delta (1 + 2(a_\mathrm{os}/w_0)^2)},
\end{align}
where $\Omega_R$ is the Rabi frequency, $\Delta$ is the detuning of the light frequency from the resonance transition,  $a_\mathrm{os}$
is the length scale of the harmonic potential that the atoms are trapped in, and $w_0$ is the beam waist of the optical mode.  For example, with $a_\mathrm{os} = 5~\mu\mathrm{m}$~\cite{julienne1997}, $w_0 = 30~\mu\mathrm{m}$, $\Delta = 2\pi \times 100~\mathrm{MHz}$, saturation intensity $I_\mathrm{sat} = 17~\mathrm{W/m^2}$~\cite{steck2021}, atomic decay rate $\gamma = 2\pi \times  1~\mathrm{MHz}$, incident power of the laser $P = 7~\mathrm{mW}$ gives intensity at the beam waist $I = 5\mathrm{MW}$, and $\Omega_R^2 = I \gamma^2/(2I_\mathrm{sat}) $ gives the coupling strength $\Omega = 2\pi \times 1.4~\mathrm{GHz}$. Such coupling strengths are easily attainable with the current state-of-the-art experiments. Larger couplings can be obtained by reducing the detuning or increasing the intensity. However, using the latter to increase the coupling strength may also increase the effective spontaneous emission rate that would in turn affect the coherence manipulation of the atoms. For the dephasing rate (\ref{dephasingrateformula}), using the above parameters, and varying the power of the incident the laser from $0.7~\mathrm{mW}$ to $7~\mathrm{mW}$ would give an effective decay rate $\Gamma$ in the range $1.4\times10^{-3}  < \Gamma/\Omega < 1.4\times 10^{-1}$.

\section{Conclusion}
\label{conclusion}

We have analyzed the effects of dephasing and photon loss on a scheme for creating entanglement between two BECs via a QND measurement. 
The main results that we have found are that at short time scales $ \tau < 1/\sqrt{N}$ the correlations are relatively unaffected by dephasing, but increases for larger time scales.  The origin of this is a broadening of the probability distributions, weakening the correlations and anti-correlations.  We have found entanglement survives in the presence of dephasing as long as the interaction times are in the range $ \tau < 1/\sqrt{N}$.  Beyond these times, Schrodinger cat-like stater generated which are highly sensitive to dephasing in an appropriate basis. In the presence of photon loss, we find at the QND states are highly robust against this type of decoherence, except at some particular interaction times.  Even considering cases with interaction times $ \tau > 1/\sqrt{N}$ which is beyond the Holstein-Primakoff regime, the states are remarkably robust against photon loss.  Even for the loss probabilities in the region of $95\%$ most of the correlations are still intact.  In particular, the entanglement is remarkably robust and can be observed even for large photon losses and highly non-Gaussian states. In summary, dephasing seems to be the main obstacle to be overcome in the context of generating non-Gaussian states, and photon loss is a less serious problem.  Even with dephasing, depending upon the particular type of state being targeted, the states can be robust even for large $ N $. 

We attribute this robustness to photon loss to the fact that the entanglement generation scheme only depends upon the interference of coherent states, and as long as the light amplitudes have not entirely decayed to zero, the photon loss only acts to damp the amplitudes of the coherent states.  Another way to understand the robustness of photon loss is that the decoherence factors that enter the density matrix happen to coincide with regions where the QND interaction suppresses the amplitudes.  This explains why the experimentally observed entanglement as seen in Ref. \cite{julsgaard2001experimental} could be performed without the use of any special experimental apparatus to minimize loss (e.g. cavities etc.).  Entanglement between two spatially separated BECs has never been achieved experimentally to date but our results suggest that entanglement should be achievable using the QND approach.  We have only considered two potential threats to entanglement generation, dephasing and photon loss in this paper.  While we consider these to be the most dangerous effects, there are of course other possibilities, such as atomic loss and heating of the BEC due to the QND interaction.  We leave these aspects as future work.  


\section*{Acknowledgments}
We thank Matthew Prest for providing valuable comments on the manuscript.  This work is supported by the National Natural Science Foundation of China (62071301); State Council of the People’s Republic of China (D1210036A); NSFC Research Fund for International Young Scientists (11850410426); NYU-ECNU Institute of Physics at NYU Shanghai; the Science and Technology Commission of Shanghai Municipality (19XD1423000); the China Science and Technology Exchange Center (NGA-16-001); the NYU Shanghai Boost Fund. 

\appendix

\section{Amplitude of a beam split coherent state}
\label{sec:beamsplitcoh}

In this section, we will derive the coefficients $ C_{n_c n_d} (\chi) $ shown in (\ref{cfuncdef}) corresponding to the Fock state amplitudes of a coherent state entering a beam splitter. Assuming a coherent state in mode $a$ enters a  beam splitter and is divided into two modes according to the following transformation
\begin{align}
a = c \cos \chi + d \sin \chi. 
\end{align}
Then the amplitude of the coherent state $ \alpha $ can be obtained as
\begin{align}
|\alpha \rangle_a & = e^{- | \alpha |^2/2} e^{ \alpha \cos \chi c^\dagger + \alpha \sin \chi d^\dagger} | \text{vac} \rangle \nonumber \\
& = | \alpha \cos \chi \rangle_c | \alpha \sin \chi \rangle_d \nonumber \\
& = \sum_{n_c, n_d = 0}^\infty C_{n_c n_d} (\chi) |n_c \rangle |n_d \rangle . 
\end{align}
The last line shows the claimed relation.

\section{Transformation expression between Fock states in various bases}
\label{app:trans}

The Fock eigenstates of the spin operators corresponding to $ S^x $ and $ S^y $ operators are given by
\begin{align}
| k \rangle^{(x)} & = e^{-iS^y \pi/4} | k \rangle \nonumber \\ 
| k \rangle^{(y)} & = e^{-iS^z \pi/4} e^{-iS^y \pi/4} | k \rangle. 
\end{align}
The matrix elements of the $ S^y $ rotation are given by
\begin{align}
& \langle k | e^{-i S^y \theta/2} | k' \rangle = \sqrt{ k'! (N-k')! k! (N-k)!} \nonumber \\
& \times 
\sum_n \frac{(-1)^n \cos^{ k- k' + N - 2n} (\theta/2) \sin^{2n + k' - k} (\theta/2) }{(k-n)!(N-k'-n)!n!(k'-k+n)!}, 
\label{syrotmatrixelement}
\end{align}
where $ | k \rangle $ is the eigenstates of $ S^z $. The matrix elements of $ S^x $ are accordingly given by
\begin{align}
& \langle k | e^{-i S^x \theta/2} | k' \rangle = 
i^{k'-k} \langle k | e^{-i S^y \theta/2} | k' \rangle,
\label{sxrotmatrixelement}
\end{align}
using the relation $ S^x = e^{-i S^z \pi/4} S^y e^{i S^z \pi/4} $.

\bibliographystyle{apsrev}
\bibliography{gaoshuai}

\begin{thebibliography}{106}
\expandafter\ifx\csname natexlab\endcsname\relax\def\natexlab#1{#1}\fi
\expandafter\ifx\csname bibnamefont\endcsname\relax
  \def\bibnamefont#1{#1}\fi
\expandafter\ifx\csname bibfnamefont\endcsname\relax
  \def\bibfnamefont#1{#1}\fi
\expandafter\ifx\csname citenamefont\endcsname\relax
  \def\citenamefont#1{#1}\fi
\expandafter\ifx\csname url\endcsname\relax
  \def\url#1{\texttt{#1}}\fi
\expandafter\ifx\csname urlprefix\endcsname\relax\def\urlprefix{URL }\fi
\providecommand{\bibinfo}[2]{#2}
\providecommand{\eprint}[2][]{\url{#2}}

\bibitem[{\citenamefont{Zurek}(2003)}]{zurek2003decoherence}
\bibinfo{author}{\bibfnamefont{W.~H.} \bibnamefont{Zurek}},
  \bibinfo{journal}{Reviews of modern physics} \textbf{\bibinfo{volume}{75}},
  \bibinfo{pages}{715} (\bibinfo{year}{2003}).

\bibitem[{\citenamefont{Schlosshauer}(2007)}]{schlosshauer2007decoherence}
\bibinfo{author}{\bibfnamefont{M.~A.} \bibnamefont{Schlosshauer}},
  \emph{\bibinfo{title}{Decoherence: and the quantum-to-classical transition}}
  (\bibinfo{publisher}{Springer Science \& Business Media},
  \bibinfo{year}{2007}).

\bibitem[{\citenamefont{Hornberger}(2009)}]{hornberger2009introduction}
\bibinfo{author}{\bibfnamefont{K.}~\bibnamefont{Hornberger}}, in
  \emph{\bibinfo{booktitle}{Entanglement and decoherence}}
  (\bibinfo{publisher}{Springer}, \bibinfo{year}{2009}), pp.
  \bibinfo{pages}{221--276}.

\bibitem[{\citenamefont{Lukin}(2003)}]{lukin2003colloquium}
\bibinfo{author}{\bibfnamefont{M.}~\bibnamefont{Lukin}},
  \bibinfo{journal}{Reviews of Modern Physics} \textbf{\bibinfo{volume}{75}},
  \bibinfo{pages}{457} (\bibinfo{year}{2003}).

\bibitem[{\citenamefont{Riedel et~al.}(2010)\citenamefont{Riedel, B{\"o}hi, Li,
  H{\"a}nsch, Sinatra, and Treutlein}}]{riedel2010atom}
\bibinfo{author}{\bibfnamefont{M.~F.} \bibnamefont{Riedel}},
  \bibinfo{author}{\bibfnamefont{P.}~\bibnamefont{B{\"o}hi}},
  \bibinfo{author}{\bibfnamefont{Y.}~\bibnamefont{Li}},
  \bibinfo{author}{\bibfnamefont{T.~W.} \bibnamefont{H{\"a}nsch}},
  \bibinfo{author}{\bibfnamefont{A.}~\bibnamefont{Sinatra}}, \bibnamefont{and}
  \bibinfo{author}{\bibfnamefont{P.}~\bibnamefont{Treutlein}},
  \bibinfo{journal}{Nature} \textbf{\bibinfo{volume}{464}},
  \bibinfo{pages}{1170} (\bibinfo{year}{2010}).

\bibitem[{\citenamefont{Pezze et~al.}(2018)\citenamefont{Pezze, Smerzi,
  Oberthaler, Schmied, and Treutlein}}]{pezze2018quantum}
\bibinfo{author}{\bibfnamefont{L.}~\bibnamefont{Pezze}},
  \bibinfo{author}{\bibfnamefont{A.}~\bibnamefont{Smerzi}},
  \bibinfo{author}{\bibfnamefont{M.~K.} \bibnamefont{Oberthaler}},
  \bibinfo{author}{\bibfnamefont{R.}~\bibnamefont{Schmied}}, \bibnamefont{and}
  \bibinfo{author}{\bibfnamefont{P.}~\bibnamefont{Treutlein}},
  \bibinfo{journal}{Reviews of Modern Physics} \textbf{\bibinfo{volume}{90}},
  \bibinfo{pages}{035005} (\bibinfo{year}{2018}).

\bibitem[{\citenamefont{O’Connell et~al.}(2010)\citenamefont{O’Connell,
  Hofheinz, Ansmann, Bialczak, Lenander, Lucero, Neeley, Sank, Wang, Weides
  et~al.}}]{o2010quantum}
\bibinfo{author}{\bibfnamefont{A.~D.} \bibnamefont{O’Connell}},
  \bibinfo{author}{\bibfnamefont{M.}~\bibnamefont{Hofheinz}},
  \bibinfo{author}{\bibfnamefont{M.}~\bibnamefont{Ansmann}},
  \bibinfo{author}{\bibfnamefont{R.~C.} \bibnamefont{Bialczak}},
  \bibinfo{author}{\bibfnamefont{M.}~\bibnamefont{Lenander}},
  \bibinfo{author}{\bibfnamefont{E.}~\bibnamefont{Lucero}},
  \bibinfo{author}{\bibfnamefont{M.}~\bibnamefont{Neeley}},
  \bibinfo{author}{\bibfnamefont{D.}~\bibnamefont{Sank}},
  \bibinfo{author}{\bibfnamefont{H.}~\bibnamefont{Wang}},
  \bibinfo{author}{\bibfnamefont{M.}~\bibnamefont{Weides}},
  \bibnamefont{et~al.}, \bibinfo{journal}{Nature}
  \textbf{\bibinfo{volume}{464}}, \bibinfo{pages}{697} (\bibinfo{year}{2010}).

\bibitem[{\citenamefont{Sørensen et~al.}(2001)\citenamefont{Sørensen, Duan,
  Cirac, and Zoller}}]{sorensen2001many}
\bibinfo{author}{\bibfnamefont{A.}~\bibnamefont{Sørensen}},
  \bibinfo{author}{\bibfnamefont{L.-M.} \bibnamefont{Duan}},
  \bibinfo{author}{\bibfnamefont{J.~I.} \bibnamefont{Cirac}}, \bibnamefont{and}
  \bibinfo{author}{\bibfnamefont{P.}~\bibnamefont{Zoller}},
  \bibinfo{journal}{Nature} \textbf{\bibinfo{volume}{409}},
  \bibinfo{pages}{63–66} (\bibinfo{year}{2001}), ISSN
  \bibinfo{issn}{1476-4687}.

\bibitem[{\citenamefont{Machida et~al.}(1987)\citenamefont{Machida, Yamamoto,
  and Itaya}}]{machida1987observation}
\bibinfo{author}{\bibfnamefont{S.}~\bibnamefont{Machida}},
  \bibinfo{author}{\bibfnamefont{Y.}~\bibnamefont{Yamamoto}}, \bibnamefont{and}
  \bibinfo{author}{\bibfnamefont{Y.}~\bibnamefont{Itaya}},
  \bibinfo{journal}{Physical review letters} \textbf{\bibinfo{volume}{58}},
  \bibinfo{pages}{1000} (\bibinfo{year}{1987}).

\bibitem[{\citenamefont{Shelby et~al.}(1986)\citenamefont{Shelby, Levenson,
  Perlmutter, DeVoe, and Walls}}]{shelby1986broad}
\bibinfo{author}{\bibfnamefont{R.}~\bibnamefont{Shelby}},
  \bibinfo{author}{\bibfnamefont{M.}~\bibnamefont{Levenson}},
  \bibinfo{author}{\bibfnamefont{S.}~\bibnamefont{Perlmutter}},
  \bibinfo{author}{\bibfnamefont{R.}~\bibnamefont{DeVoe}}, \bibnamefont{and}
  \bibinfo{author}{\bibfnamefont{D.}~\bibnamefont{Walls}},
  \bibinfo{journal}{Physical review letters} \textbf{\bibinfo{volume}{57}},
  \bibinfo{pages}{691} (\bibinfo{year}{1986}).

\bibitem[{\citenamefont{Slusher et~al.}(1986)\citenamefont{Slusher, Hollberg,
  Yurke, Mertz, and Valley}}]{slusher1986observation}
\bibinfo{author}{\bibfnamefont{R.}~\bibnamefont{Slusher}},
  \bibinfo{author}{\bibfnamefont{L.}~\bibnamefont{Hollberg}},
  \bibinfo{author}{\bibfnamefont{B.}~\bibnamefont{Yurke}},
  \bibinfo{author}{\bibfnamefont{J.}~\bibnamefont{Mertz}}, \bibnamefont{and}
  \bibinfo{author}{\bibfnamefont{J.}~\bibnamefont{Valley}},
  \bibinfo{journal}{Physical Review Letters} \textbf{\bibinfo{volume}{56}},
  \bibinfo{pages}{788} (\bibinfo{year}{1986}).

\bibitem[{\citenamefont{Wu et~al.}(1986)\citenamefont{Wu, Kimble, Hall, and
  Wu}}]{wu1986generation}
\bibinfo{author}{\bibfnamefont{L.-A.} \bibnamefont{Wu}},
  \bibinfo{author}{\bibfnamefont{H.}~\bibnamefont{Kimble}},
  \bibinfo{author}{\bibfnamefont{J.}~\bibnamefont{Hall}}, \bibnamefont{and}
  \bibinfo{author}{\bibfnamefont{H.}~\bibnamefont{Wu}},
  \bibinfo{journal}{Physical review letters} \textbf{\bibinfo{volume}{57}},
  \bibinfo{pages}{2520} (\bibinfo{year}{1986}).

\bibitem[{\citenamefont{Schwarzhans et~al.}(1997)\citenamefont{Schwarzhans,
  Rauch, and Weimerkirch}}]{Schwarzhans1997}
\bibinfo{author}{\bibfnamefont{M.}~\bibnamefont{Schwarzhans}},
  \bibinfo{author}{\bibfnamefont{L.}~\bibnamefont{Rauch}}, \bibnamefont{and}
  \bibinfo{author}{\bibfnamefont{M.~J.~J.} \bibnamefont{Weimerkirch}},
  \bibinfo{journal}{Nature} \textbf{\bibinfo{volume}{387}},
  \bibinfo{pages}{471} (\bibinfo{year}{1997}).

\bibitem[{\citenamefont{Slusher et~al.}(1985)\citenamefont{Slusher, Hollberg,
  Yurke, Mertz, and Valley}}]{slusher1985observation}
\bibinfo{author}{\bibfnamefont{R.}~\bibnamefont{Slusher}},
  \bibinfo{author}{\bibfnamefont{L.}~\bibnamefont{Hollberg}},
  \bibinfo{author}{\bibfnamefont{B.}~\bibnamefont{Yurke}},
  \bibinfo{author}{\bibfnamefont{J.}~\bibnamefont{Mertz}}, \bibnamefont{and}
  \bibinfo{author}{\bibfnamefont{J.}~\bibnamefont{Valley}},
  \bibinfo{journal}{Physical Review Letters} \textbf{\bibinfo{volume}{55}},
  \bibinfo{pages}{2409} (\bibinfo{year}{1985}).

\bibitem[{\citenamefont{Breitenbach et~al.}(1997)\citenamefont{Breitenbach,
  Schiller, and Mlynek}}]{breitenbach1997measurement}
\bibinfo{author}{\bibfnamefont{G.}~\bibnamefont{Breitenbach}},
  \bibinfo{author}{\bibfnamefont{S.}~\bibnamefont{Schiller}}, \bibnamefont{and}
  \bibinfo{author}{\bibfnamefont{J.}~\bibnamefont{Mlynek}},
  \bibinfo{journal}{Nature} \textbf{\bibinfo{volume}{387}},
  \bibinfo{pages}{471} (\bibinfo{year}{1997}).

\bibitem[{\citenamefont{Gross}(2012)}]{gross2012spin}
\bibinfo{author}{\bibfnamefont{C.}~\bibnamefont{Gross}},
  \bibinfo{journal}{Journal of Physics B: Atomic, Molecular and Optical
  Physics} \textbf{\bibinfo{volume}{45}}, \bibinfo{pages}{103001}
  (\bibinfo{year}{2012}).

\bibitem[{\citenamefont{Byrnes and Ilo-Okeke}(2021)}]{byrnes2020quantum}
\bibinfo{author}{\bibfnamefont{T.}~\bibnamefont{Byrnes}} \bibnamefont{and}
  \bibinfo{author}{\bibfnamefont{E.~O.} \bibnamefont{Ilo-Okeke}},
  \emph{\bibinfo{title}{Quantum atom optics: Theory and applications to quantum
  technology}} (\bibinfo{publisher}{Cambridge university press},
  \bibinfo{year}{2021}).

\bibitem[{\citenamefont{Esteve et~al.}(2008)\citenamefont{Esteve, Gross,
  Weller, Giovanazzi, and Oberthaler}}]{esteve2008squeezing}
\bibinfo{author}{\bibfnamefont{J.}~\bibnamefont{Esteve}},
  \bibinfo{author}{\bibfnamefont{C.}~\bibnamefont{Gross}},
  \bibinfo{author}{\bibfnamefont{A.}~\bibnamefont{Weller}},
  \bibinfo{author}{\bibfnamefont{S.}~\bibnamefont{Giovanazzi}},
  \bibnamefont{and}
  \bibinfo{author}{\bibfnamefont{M.}~\bibnamefont{Oberthaler}},
  \bibinfo{journal}{Nature} \textbf{\bibinfo{volume}{455}},
  \bibinfo{pages}{1216} (\bibinfo{year}{2008}).

\bibitem[{\citenamefont{Macomber and Lynch}(1985)}]{Macomber1985a}
\bibinfo{author}{\bibfnamefont{J.~D.} \bibnamefont{Macomber}} \bibnamefont{and}
  \bibinfo{author}{\bibfnamefont{R.}~\bibnamefont{Lynch}},
  \bibinfo{journal}{The Journal of Chemical Physics}
  \textbf{\bibinfo{volume}{83}}, \bibinfo{pages}{6514} (\bibinfo{year}{1985}),
  ISSN \bibinfo{issn}{00219606}.

\bibitem[{\citenamefont{Walls and Zoller}(1981)}]{PhysRevLett.47.709}
\bibinfo{author}{\bibfnamefont{D.~F.} \bibnamefont{Walls}} \bibnamefont{and}
  \bibinfo{author}{\bibfnamefont{P.}~\bibnamefont{Zoller}},
  \bibinfo{journal}{Phys. Rev. Lett.} \textbf{\bibinfo{volume}{47}},
  \bibinfo{pages}{709} (\bibinfo{year}{1981}).

\bibitem[{\citenamefont{Wodkiewicz and Eberly}(1985)}]{Wodkiewicz:85}
\bibinfo{author}{\bibfnamefont{K.}~\bibnamefont{Wodkiewicz}} \bibnamefont{and}
  \bibinfo{author}{\bibfnamefont{J.~H.} \bibnamefont{Eberly}},
  \bibinfo{journal}{J. Opt. Soc. Am. B} \textbf{\bibinfo{volume}{2}},
  \bibinfo{pages}{458} (\bibinfo{year}{1985}).

\bibitem[{\citenamefont{Wineland et~al.}(1992)\citenamefont{Wineland,
  Bollinger, Itano, Moore, and Heinzen}}]{PhysRevA.46.R6797}
\bibinfo{author}{\bibfnamefont{D.~J.} \bibnamefont{Wineland}},
  \bibinfo{author}{\bibfnamefont{J.~J.} \bibnamefont{Bollinger}},
  \bibinfo{author}{\bibfnamefont{W.~M.} \bibnamefont{Itano}},
  \bibinfo{author}{\bibfnamefont{F.~L.} \bibnamefont{Moore}}, \bibnamefont{and}
  \bibinfo{author}{\bibfnamefont{D.~J.} \bibnamefont{Heinzen}},
  \bibinfo{journal}{Phys. Rev. A} \textbf{\bibinfo{volume}{46}},
  \bibinfo{pages}{R6797} (\bibinfo{year}{1992}).

\bibitem[{\citenamefont{Vernac et~al.}(2000)\citenamefont{Vernac, Pinard, and
  Giacobino}}]{vernac2000spin}
\bibinfo{author}{\bibfnamefont{L.}~\bibnamefont{Vernac}},
  \bibinfo{author}{\bibfnamefont{M.}~\bibnamefont{Pinard}}, \bibnamefont{and}
  \bibinfo{author}{\bibfnamefont{E.}~\bibnamefont{Giacobino}},
  \bibinfo{journal}{Physical Review A} \textbf{\bibinfo{volume}{62}},
  \bibinfo{pages}{063812} (\bibinfo{year}{2000}).

\bibitem[{\citenamefont{Zhang and Duan}(2014)}]{zhang2014quantum}
\bibinfo{author}{\bibfnamefont{Z.}~\bibnamefont{Zhang}} \bibnamefont{and}
  \bibinfo{author}{\bibfnamefont{L.}~\bibnamefont{Duan}}, \bibinfo{journal}{New
  Journal of Physics} \textbf{\bibinfo{volume}{16}}, \bibinfo{pages}{103037}
  (\bibinfo{year}{2014}).

\bibitem[{\citenamefont{Kitagawa and Ueda}(1993)}]{kitagawa1993squeezed}
\bibinfo{author}{\bibfnamefont{M.}~\bibnamefont{Kitagawa}} \bibnamefont{and}
  \bibinfo{author}{\bibfnamefont{M.}~\bibnamefont{Ueda}},
  \bibinfo{journal}{Physical Review A} \textbf{\bibinfo{volume}{47}},
  \bibinfo{pages}{5138} (\bibinfo{year}{1993}).

\bibitem[{\citenamefont{Hald et~al.}(1999)\citenamefont{Hald, S{\o}rensen,
  Schori, and Polzik}}]{hald1999spin}
\bibinfo{author}{\bibfnamefont{J.}~\bibnamefont{Hald}},
  \bibinfo{author}{\bibfnamefont{J.}~\bibnamefont{S{\o}rensen}},
  \bibinfo{author}{\bibfnamefont{C.}~\bibnamefont{Schori}}, \bibnamefont{and}
  \bibinfo{author}{\bibfnamefont{E.}~\bibnamefont{Polzik}},
  \bibinfo{journal}{Physical review letters} \textbf{\bibinfo{volume}{83}},
  \bibinfo{pages}{1319} (\bibinfo{year}{1999}).

\bibitem[{\citenamefont{Orzel et~al.}(2001)\citenamefont{Orzel, Tuchman,
  Fenselau, Yasuda, and Kasevich}}]{orzel2001squeezed}
\bibinfo{author}{\bibfnamefont{C.}~\bibnamefont{Orzel}},
  \bibinfo{author}{\bibfnamefont{A.}~\bibnamefont{Tuchman}},
  \bibinfo{author}{\bibfnamefont{M.}~\bibnamefont{Fenselau}},
  \bibinfo{author}{\bibfnamefont{M.}~\bibnamefont{Yasuda}}, \bibnamefont{and}
  \bibinfo{author}{\bibfnamefont{M.}~\bibnamefont{Kasevich}},
  \bibinfo{journal}{Science} \textbf{\bibinfo{volume}{291}},
  \bibinfo{pages}{2386} (\bibinfo{year}{2001}).

\bibitem[{\citenamefont{Jo et~al.}(2007)\citenamefont{Jo, Shin, Will, Pasquini,
  Saba, Ketterle, Pritchard, Vengalattore, and Prentiss}}]{jo2007long}
\bibinfo{author}{\bibfnamefont{G.-B.} \bibnamefont{Jo}},
  \bibinfo{author}{\bibfnamefont{Y.}~\bibnamefont{Shin}},
  \bibinfo{author}{\bibfnamefont{S.}~\bibnamefont{Will}},
  \bibinfo{author}{\bibfnamefont{T.}~\bibnamefont{Pasquini}},
  \bibinfo{author}{\bibfnamefont{M.}~\bibnamefont{Saba}},
  \bibinfo{author}{\bibfnamefont{W.}~\bibnamefont{Ketterle}},
  \bibinfo{author}{\bibfnamefont{D.~E.} \bibnamefont{Pritchard}},
  \bibinfo{author}{\bibfnamefont{M.}~\bibnamefont{Vengalattore}},
  \bibnamefont{and} \bibinfo{author}{\bibfnamefont{M.}~\bibnamefont{Prentiss}},
  \bibinfo{journal}{Physical Review Letters} \textbf{\bibinfo{volume}{98}},
  \bibinfo{pages}{030407} (\bibinfo{year}{2007}).

\bibitem[{\citenamefont{B{\"o}hi and Riedel}(2009)}]{bohi2009j}
\bibinfo{author}{\bibfnamefont{M.}~\bibnamefont{B{\"o}hi}} \bibnamefont{and}
  \bibinfo{author}{\bibfnamefont{J.~H.} \bibnamefont{Riedel}},
  \bibinfo{journal}{Nat. Phys} \textbf{\bibinfo{volume}{5}},
  \bibinfo{pages}{592} (\bibinfo{year}{2009}).

\bibitem[{\citenamefont{Krauter et~al.}(2011)\citenamefont{Krauter, Muschik,
  Jensen, Wasilewski, Petersen, Cirac, and Polzik}}]{krauter2011entanglement}
\bibinfo{author}{\bibfnamefont{H.}~\bibnamefont{Krauter}},
  \bibinfo{author}{\bibfnamefont{C.~A.} \bibnamefont{Muschik}},
  \bibinfo{author}{\bibfnamefont{K.}~\bibnamefont{Jensen}},
  \bibinfo{author}{\bibfnamefont{W.}~\bibnamefont{Wasilewski}},
  \bibinfo{author}{\bibfnamefont{J.~M.} \bibnamefont{Petersen}},
  \bibinfo{author}{\bibfnamefont{J.~I.} \bibnamefont{Cirac}}, \bibnamefont{and}
  \bibinfo{author}{\bibfnamefont{E.~S.} \bibnamefont{Polzik}},
  \bibinfo{journal}{Physical review letters} \textbf{\bibinfo{volume}{107}},
  \bibinfo{pages}{080503} (\bibinfo{year}{2011}).

\bibitem[{\citenamefont{Riedel}(2010)}]{riedel2010p}
\bibinfo{author}{\bibfnamefont{M.}~\bibnamefont{Riedel}},
  \bibinfo{journal}{Nature} \textbf{\bibinfo{volume}{464}},
  \bibinfo{pages}{08988} (\bibinfo{year}{2010}).

\bibitem[{\citenamefont{Muessel et~al.}(2014)\citenamefont{Muessel, Strobel,
  Linnemann, Hume, and Oberthaler}}]{muessel2014scalable}
\bibinfo{author}{\bibfnamefont{W.}~\bibnamefont{Muessel}},
  \bibinfo{author}{\bibfnamefont{H.}~\bibnamefont{Strobel}},
  \bibinfo{author}{\bibfnamefont{D.}~\bibnamefont{Linnemann}},
  \bibinfo{author}{\bibfnamefont{D.}~\bibnamefont{Hume}}, \bibnamefont{and}
  \bibinfo{author}{\bibfnamefont{M.}~\bibnamefont{Oberthaler}},
  \bibinfo{journal}{Physical Review Letters} \textbf{\bibinfo{volume}{113}},
  \bibinfo{pages}{103004} (\bibinfo{year}{2014}).

\bibitem[{\citenamefont{Bao et~al.}(2020{\natexlab{a}})\citenamefont{Bao, Duan,
  Jin, Lu, Li, Qu, Wang, Novikova, Mikhailov, Zhao et~al.}}]{bao2020spin}
\bibinfo{author}{\bibfnamefont{H.}~\bibnamefont{Bao}},
  \bibinfo{author}{\bibfnamefont{J.}~\bibnamefont{Duan}},
  \bibinfo{author}{\bibfnamefont{S.}~\bibnamefont{Jin}},
  \bibinfo{author}{\bibfnamefont{X.}~\bibnamefont{Lu}},
  \bibinfo{author}{\bibfnamefont{P.}~\bibnamefont{Li}},
  \bibinfo{author}{\bibfnamefont{W.}~\bibnamefont{Qu}},
  \bibinfo{author}{\bibfnamefont{M.}~\bibnamefont{Wang}},
  \bibinfo{author}{\bibfnamefont{I.}~\bibnamefont{Novikova}},
  \bibinfo{author}{\bibfnamefont{E.~E.} \bibnamefont{Mikhailov}},
  \bibinfo{author}{\bibfnamefont{K.-F.} \bibnamefont{Zhao}},
  \bibnamefont{et~al.}, \bibinfo{journal}{Nature}
  \textbf{\bibinfo{volume}{581}}, \bibinfo{pages}{159}
  (\bibinfo{year}{2020}{\natexlab{a}}).

\bibitem[{\citenamefont{Bao et~al.}(2020{\natexlab{b}})\citenamefont{Bao, Jin,
  Duan, Jia, M{\o}lmer, Shen, and Xiao}}]{bao2020retrodiction}
\bibinfo{author}{\bibfnamefont{H.}~\bibnamefont{Bao}},
  \bibinfo{author}{\bibfnamefont{S.}~\bibnamefont{Jin}},
  \bibinfo{author}{\bibfnamefont{J.}~\bibnamefont{Duan}},
  \bibinfo{author}{\bibfnamefont{S.}~\bibnamefont{Jia}},
  \bibinfo{author}{\bibfnamefont{K.}~\bibnamefont{M{\o}lmer}},
  \bibinfo{author}{\bibfnamefont{H.}~\bibnamefont{Shen}}, \bibnamefont{and}
  \bibinfo{author}{\bibfnamefont{Y.}~\bibnamefont{Xiao}},
  \bibinfo{journal}{Nature communications} \textbf{\bibinfo{volume}{11}},
  \bibinfo{pages}{1} (\bibinfo{year}{2020}{\natexlab{b}}).

\bibitem[{\citenamefont{Hammerer et~al.}(2010)\citenamefont{Hammerer,
  S{\o}rensen, and Polzik}}]{hammerer2010quantum}
\bibinfo{author}{\bibfnamefont{K.}~\bibnamefont{Hammerer}},
  \bibinfo{author}{\bibfnamefont{A.~S.} \bibnamefont{S{\o}rensen}},
  \bibnamefont{and} \bibinfo{author}{\bibfnamefont{E.~S.}
  \bibnamefont{Polzik}}, \bibinfo{journal}{Reviews of Modern Physics}
  \textbf{\bibinfo{volume}{82}}, \bibinfo{pages}{1041} (\bibinfo{year}{2010}).

\bibitem[{\citenamefont{Hammerer et~al.}(2004)\citenamefont{Hammerer,
  M{\o}lmer, Polzik, and Cirac}}]{hammerer2004light}
\bibinfo{author}{\bibfnamefont{K.}~\bibnamefont{Hammerer}},
  \bibinfo{author}{\bibfnamefont{K.}~\bibnamefont{M{\o}lmer}},
  \bibinfo{author}{\bibfnamefont{E.~S.} \bibnamefont{Polzik}},
  \bibnamefont{and} \bibinfo{author}{\bibfnamefont{J.~I.} \bibnamefont{Cirac}},
  \bibinfo{journal}{Physical Review A} \textbf{\bibinfo{volume}{70}},
  \bibinfo{pages}{044304} (\bibinfo{year}{2004}).

\bibitem[{\citenamefont{Appel et~al.}(2009)\citenamefont{Appel, Windpassinger,
  Oblak, Hoff, Kj{\ae}rgaard, and Polzik}}]{appel2009mesoscopic}
\bibinfo{author}{\bibfnamefont{J.}~\bibnamefont{Appel}},
  \bibinfo{author}{\bibfnamefont{P.~J.} \bibnamefont{Windpassinger}},
  \bibinfo{author}{\bibfnamefont{D.}~\bibnamefont{Oblak}},
  \bibinfo{author}{\bibfnamefont{U.~B.} \bibnamefont{Hoff}},
  \bibinfo{author}{\bibfnamefont{N.}~\bibnamefont{Kj{\ae}rgaard}},
  \bibnamefont{and} \bibinfo{author}{\bibfnamefont{E.~S.}
  \bibnamefont{Polzik}}, \bibinfo{journal}{Proceedings of the National Academy
  of Sciences} \textbf{\bibinfo{volume}{106}}, \bibinfo{pages}{10960}
  (\bibinfo{year}{2009}).

\bibitem[{\citenamefont{Eckert et~al.}(2008)\citenamefont{Eckert, Romero-Isart,
  Rodriguez, Lewenstein, Polzik, and Sanpera}}]{eckert2008quantum}
\bibinfo{author}{\bibfnamefont{K.}~\bibnamefont{Eckert}},
  \bibinfo{author}{\bibfnamefont{O.}~\bibnamefont{Romero-Isart}},
  \bibinfo{author}{\bibfnamefont{M.}~\bibnamefont{Rodriguez}},
  \bibinfo{author}{\bibfnamefont{M.}~\bibnamefont{Lewenstein}},
  \bibinfo{author}{\bibfnamefont{E.~S.} \bibnamefont{Polzik}},
  \bibnamefont{and} \bibinfo{author}{\bibfnamefont{A.}~\bibnamefont{Sanpera}},
  \bibinfo{journal}{Nature Physics} \textbf{\bibinfo{volume}{4}},
  \bibinfo{pages}{50} (\bibinfo{year}{2008}).

\bibitem[{\citenamefont{Sewell et~al.}(2012)\citenamefont{Sewell, Koschorreck,
  Napolitano, Dubost, Behbood, and Mitchell}}]{sewell2012magnetic}
\bibinfo{author}{\bibfnamefont{R.}~\bibnamefont{Sewell}},
  \bibinfo{author}{\bibfnamefont{M.}~\bibnamefont{Koschorreck}},
  \bibinfo{author}{\bibfnamefont{M.}~\bibnamefont{Napolitano}},
  \bibinfo{author}{\bibfnamefont{B.}~\bibnamefont{Dubost}},
  \bibinfo{author}{\bibfnamefont{N.}~\bibnamefont{Behbood}}, \bibnamefont{and}
  \bibinfo{author}{\bibfnamefont{M.}~\bibnamefont{Mitchell}},
  \bibinfo{journal}{Physical review letters} \textbf{\bibinfo{volume}{109}},
  \bibinfo{pages}{253605} (\bibinfo{year}{2012}).

\bibitem[{\citenamefont{Koschorreck et~al.}(2010)\citenamefont{Koschorreck,
  Napolitano, Dubost, and Mitchell}}]{koschorreck2010sub}
\bibinfo{author}{\bibfnamefont{M.}~\bibnamefont{Koschorreck}},
  \bibinfo{author}{\bibfnamefont{M.}~\bibnamefont{Napolitano}},
  \bibinfo{author}{\bibfnamefont{B.}~\bibnamefont{Dubost}}, \bibnamefont{and}
  \bibinfo{author}{\bibfnamefont{M.}~\bibnamefont{Mitchell}},
  \bibinfo{journal}{Physical review letters} \textbf{\bibinfo{volume}{104}},
  \bibinfo{pages}{093602} (\bibinfo{year}{2010}).

\bibitem[{\citenamefont{Colangelo et~al.}(2017)\citenamefont{Colangelo,
  Ciurana, Bianchet, Sewell, and Mitchell}}]{colangelo2017simultaneous}
\bibinfo{author}{\bibfnamefont{G.}~\bibnamefont{Colangelo}},
  \bibinfo{author}{\bibfnamefont{F.~M.} \bibnamefont{Ciurana}},
  \bibinfo{author}{\bibfnamefont{L.~C.} \bibnamefont{Bianchet}},
  \bibinfo{author}{\bibfnamefont{R.~J.} \bibnamefont{Sewell}},
  \bibnamefont{and} \bibinfo{author}{\bibfnamefont{M.~W.}
  \bibnamefont{Mitchell}}, \bibinfo{journal}{Nature}
  \textbf{\bibinfo{volume}{543}}, \bibinfo{pages}{525} (\bibinfo{year}{2017}).

\bibitem[{\citenamefont{Ilo-Okeke et~al.}(2021)\citenamefont{Ilo-Okeke, Sunami,
  Foot, and Byrnes}}]{bec1}
\bibinfo{author}{\bibfnamefont{E.~O.} \bibnamefont{Ilo-Okeke}},
  \bibinfo{author}{\bibfnamefont{S.}~\bibnamefont{Sunami}},
  \bibinfo{author}{\bibfnamefont{C.~J.} \bibnamefont{Foot}}, \bibnamefont{and}
  \bibinfo{author}{\bibfnamefont{T.}~\bibnamefont{Byrnes}},
  \emph{\bibinfo{title}{Faraday imaging induced squeezing of a double-well
  bose-einstein condensate}} (\bibinfo{year}{2021}), \eprint{2104.02382}.

\bibitem[{\citenamefont{Giovannetti et~al.}(2011)\citenamefont{Giovannetti,
  Lloyd, and Maccone}}]{giovannetti2011advances}
\bibinfo{author}{\bibfnamefont{V.}~\bibnamefont{Giovannetti}},
  \bibinfo{author}{\bibfnamefont{S.}~\bibnamefont{Lloyd}}, \bibnamefont{and}
  \bibinfo{author}{\bibfnamefont{L.}~\bibnamefont{Maccone}},
  \bibinfo{journal}{Nature photonics} \textbf{\bibinfo{volume}{5}},
  \bibinfo{pages}{222} (\bibinfo{year}{2011}).

\bibitem[{\citenamefont{You et~al.}(2017)\citenamefont{You, Adhikari, Chi,
  LaBorde, Matyas, Zhang, Su, Byrnes, Lu, Dowling
  et~al.}}]{you2017multiparameter}
\bibinfo{author}{\bibfnamefont{C.}~\bibnamefont{You}},
  \bibinfo{author}{\bibfnamefont{S.}~\bibnamefont{Adhikari}},
  \bibinfo{author}{\bibfnamefont{Y.}~\bibnamefont{Chi}},
  \bibinfo{author}{\bibfnamefont{M.~L.} \bibnamefont{LaBorde}},
  \bibinfo{author}{\bibfnamefont{C.~T.} \bibnamefont{Matyas}},
  \bibinfo{author}{\bibfnamefont{C.}~\bibnamefont{Zhang}},
  \bibinfo{author}{\bibfnamefont{Z.}~\bibnamefont{Su}},
  \bibinfo{author}{\bibfnamefont{T.}~\bibnamefont{Byrnes}},
  \bibinfo{author}{\bibfnamefont{C.}~\bibnamefont{Lu}},
  \bibinfo{author}{\bibfnamefont{J.~P.} \bibnamefont{Dowling}},
  \bibnamefont{et~al.}, \bibinfo{journal}{Journal of Optics}
  \textbf{\bibinfo{volume}{19}}, \bibinfo{pages}{124002}
  (\bibinfo{year}{2017}).

\bibitem[{\citenamefont{Bondurant and Shapiro}(1984)}]{bondurant1984squeezed}
\bibinfo{author}{\bibfnamefont{R.~S.} \bibnamefont{Bondurant}}
  \bibnamefont{and} \bibinfo{author}{\bibfnamefont{J.~H.}
  \bibnamefont{Shapiro}}, \bibinfo{journal}{Physical Review D}
  \textbf{\bibinfo{volume}{30}}, \bibinfo{pages}{2548} (\bibinfo{year}{1984}).

\bibitem[{\citenamefont{Schnabel et~al.}(2010)\citenamefont{Schnabel,
  Mavalvala, McClelland, and K.Lam}}]{gwavesqueeze}
\bibinfo{author}{\bibfnamefont{R.}~\bibnamefont{Schnabel}},
  \bibinfo{author}{\bibfnamefont{N.}~\bibnamefont{Mavalvala}},
  \bibinfo{author}{\bibfnamefont{D.~E.} \bibnamefont{McClelland}},
  \bibnamefont{and} \bibinfo{author}{\bibfnamefont{P.}~\bibnamefont{K.Lam}},
  \bibinfo{journal}{Nature Communications} \textbf{\bibinfo{volume}{1}}
  (\bibinfo{year}{2010}).

\bibitem[{\citenamefont{Horikiri et~al.}(2016)\citenamefont{Horikiri,
  Yamaguchi, Kamide, Matsuo, Byrnes, Ishida, L\"offler, H\"ofling, Shikano,
  Ogawa et~al.}}]{Horikiri16}
\bibinfo{author}{\bibfnamefont{T.}~\bibnamefont{Horikiri}},
  \bibinfo{author}{\bibfnamefont{M.}~\bibnamefont{Yamaguchi}},
  \bibinfo{author}{\bibfnamefont{K.}~\bibnamefont{Kamide}},
  \bibinfo{author}{\bibfnamefont{Y.}~\bibnamefont{Matsuo}},
  \bibinfo{author}{\bibfnamefont{T.}~\bibnamefont{Byrnes}},
  \bibinfo{author}{\bibfnamefont{N.}~\bibnamefont{Ishida}},
  \bibinfo{author}{\bibfnamefont{A.}~\bibnamefont{L\"offler}},
  \bibinfo{author}{\bibfnamefont{S.}~\bibnamefont{H\"ofling}},
  \bibinfo{author}{\bibfnamefont{Y.}~\bibnamefont{Shikano}},
  \bibinfo{author}{\bibfnamefont{T.}~\bibnamefont{Ogawa}},
  \bibnamefont{et~al.}, \bibinfo{journal}{Scientific reports}
  \textbf{\bibinfo{volume}{6}}, \bibinfo{pages}{25655} (\bibinfo{year}{2016}).

\bibitem[{\citenamefont{Kritsotakis et~al.}(2021)\citenamefont{Kritsotakis,
  Dunningham, and Haine}}]{PhysRevA.103.023318}
\bibinfo{author}{\bibfnamefont{M.}~\bibnamefont{Kritsotakis}},
  \bibinfo{author}{\bibfnamefont{J.~A.} \bibnamefont{Dunningham}},
  \bibnamefont{and} \bibinfo{author}{\bibfnamefont{S.~A.} \bibnamefont{Haine}},
  \bibinfo{journal}{Phys. Rev. A} \textbf{\bibinfo{volume}{103}},
  \bibinfo{pages}{023318} (\bibinfo{year}{2021}).

\bibitem[{\citenamefont{Kitzinger et~al.}(2020)\citenamefont{Kitzinger,
  Chaudhary, Kondappan, Ivannikov, and Byrnes}}]{bec4}
\bibinfo{author}{\bibfnamefont{J.}~\bibnamefont{Kitzinger}},
  \bibinfo{author}{\bibfnamefont{M.}~\bibnamefont{Chaudhary}},
  \bibinfo{author}{\bibfnamefont{M.}~\bibnamefont{Kondappan}},
  \bibinfo{author}{\bibfnamefont{V.}~\bibnamefont{Ivannikov}},
  \bibnamefont{and} \bibinfo{author}{\bibfnamefont{T.}~\bibnamefont{Byrnes}},
  \bibinfo{journal}{Phys. Rev. Research} \textbf{\bibinfo{volume}{2}},
  \bibinfo{pages}{033504} (\bibinfo{year}{2020}).

\bibitem[{\citenamefont{Jing et~al.}(2019)\citenamefont{Jing, Fadel, Ivannikov,
  and Byrnes}}]{bec6}
\bibinfo{author}{\bibfnamefont{Y.}~\bibnamefont{Jing}},
  \bibinfo{author}{\bibfnamefont{M.}~\bibnamefont{Fadel}},
  \bibinfo{author}{\bibfnamefont{V.}~\bibnamefont{Ivannikov}},
  \bibnamefont{and} \bibinfo{author}{\bibfnamefont{T.}~\bibnamefont{Byrnes}},
  \bibinfo{journal}{New J. Phys.} \textbf{\bibinfo{volume}{21}},
  \bibinfo{pages}{093038} (\bibinfo{year}{2019}).

\bibitem[{\citenamefont{Julsgaard
  et~al.}(2001{\natexlab{a}})\citenamefont{Julsgaard, Kozhekin, and
  Polzik}}]{polzikmacro}
\bibinfo{author}{\bibfnamefont{B.}~\bibnamefont{Julsgaard}},
  \bibinfo{author}{\bibfnamefont{A.}~\bibnamefont{Kozhekin}}, \bibnamefont{and}
  \bibinfo{author}{\bibfnamefont{E.}~\bibnamefont{Polzik}},
  \bibinfo{journal}{Nature} \textbf{\bibinfo{volume}{413}}
  (\bibinfo{year}{2001}{\natexlab{a}}).

\bibitem[{\citenamefont{Bao et~al.}(2012)\citenamefont{Bao, Xu, Li, Yuan, Lu,
  and Pan}}]{bao2012quantum}
\bibinfo{author}{\bibfnamefont{X.-H.} \bibnamefont{Bao}},
  \bibinfo{author}{\bibfnamefont{X.-F.} \bibnamefont{Xu}},
  \bibinfo{author}{\bibfnamefont{C.-M.} \bibnamefont{Li}},
  \bibinfo{author}{\bibfnamefont{Z.-S.} \bibnamefont{Yuan}},
  \bibinfo{author}{\bibfnamefont{C.-Y.} \bibnamefont{Lu}}, \bibnamefont{and}
  \bibinfo{author}{\bibfnamefont{J.-W.} \bibnamefont{Pan}},
  \bibinfo{journal}{Proceedings of the National Academy of Sciences}
  \textbf{\bibinfo{volume}{109}}, \bibinfo{pages}{20347}
  (\bibinfo{year}{2012}).

\bibitem[{\citenamefont{Krauter et~al.}(2013)\citenamefont{Krauter, Salart,
  Muschik, Petersen, Shen, Fernholz, and Polzik}}]{Krauter_2013}
\bibinfo{author}{\bibfnamefont{H.}~\bibnamefont{Krauter}},
  \bibinfo{author}{\bibfnamefont{D.}~\bibnamefont{Salart}},
  \bibinfo{author}{\bibfnamefont{C.~A.} \bibnamefont{Muschik}},
  \bibinfo{author}{\bibfnamefont{J.~M.} \bibnamefont{Petersen}},
  \bibinfo{author}{\bibfnamefont{H.}~\bibnamefont{Shen}},
  \bibinfo{author}{\bibfnamefont{T.}~\bibnamefont{Fernholz}}, \bibnamefont{and}
  \bibinfo{author}{\bibfnamefont{E.~S.} \bibnamefont{Polzik}},
  \bibinfo{journal}{Nature Physics} \textbf{\bibinfo{volume}{9}},
  \bibinfo{pages}{400–404} (\bibinfo{year}{2013}), ISSN
  \bibinfo{issn}{1745-2481}.

\bibitem[{\citenamefont{Kurkjian et~al.}(2013)\citenamefont{Kurkjian,
  Paw{\l}owski, Sinatra, and Treutlein}}]{kurkjian2013spin}
\bibinfo{author}{\bibfnamefont{H.}~\bibnamefont{Kurkjian}},
  \bibinfo{author}{\bibfnamefont{K.}~\bibnamefont{Paw{\l}owski}},
  \bibinfo{author}{\bibfnamefont{A.}~\bibnamefont{Sinatra}}, \bibnamefont{and}
  \bibinfo{author}{\bibfnamefont{P.}~\bibnamefont{Treutlein}},
  \bibinfo{journal}{Physical Review A} \textbf{\bibinfo{volume}{88}},
  \bibinfo{pages}{043605} (\bibinfo{year}{2013}).

\bibitem[{\citenamefont{Li et~al.}(2009)\citenamefont{Li, Treutlein, Reichel,
  and Sinatra}}]{li2009spin}
\bibinfo{author}{\bibfnamefont{Y.}~\bibnamefont{Li}},
  \bibinfo{author}{\bibfnamefont{P.}~\bibnamefont{Treutlein}},
  \bibinfo{author}{\bibfnamefont{J.}~\bibnamefont{Reichel}}, \bibnamefont{and}
  \bibinfo{author}{\bibfnamefont{A.}~\bibnamefont{Sinatra}},
  \bibinfo{journal}{The European Physical Journal B}
  \textbf{\bibinfo{volume}{68}}, \bibinfo{pages}{365} (\bibinfo{year}{2009}).

\bibitem[{\citenamefont{Byrnes}(2013)}]{byrnes2013fractality}
\bibinfo{author}{\bibfnamefont{T.}~\bibnamefont{Byrnes}},
  \bibinfo{journal}{Physical Review A} \textbf{\bibinfo{volume}{88}},
  \bibinfo{pages}{023609} (\bibinfo{year}{2013}).

\bibitem[{\citenamefont{Lange et~al.}(2018)\citenamefont{Lange, Peise,
  L{\"u}cke, Kruse, Vitagliano, Apellaniz, Kleinmann, T{\'o}th, and
  Klempt}}]{lange2018entanglement}
\bibinfo{author}{\bibfnamefont{K.}~\bibnamefont{Lange}},
  \bibinfo{author}{\bibfnamefont{J.}~\bibnamefont{Peise}},
  \bibinfo{author}{\bibfnamefont{B.}~\bibnamefont{L{\"u}cke}},
  \bibinfo{author}{\bibfnamefont{I.}~\bibnamefont{Kruse}},
  \bibinfo{author}{\bibfnamefont{G.}~\bibnamefont{Vitagliano}},
  \bibinfo{author}{\bibfnamefont{I.}~\bibnamefont{Apellaniz}},
  \bibinfo{author}{\bibfnamefont{M.}~\bibnamefont{Kleinmann}},
  \bibinfo{author}{\bibfnamefont{G.}~\bibnamefont{T{\'o}th}}, \bibnamefont{and}
  \bibinfo{author}{\bibfnamefont{C.}~\bibnamefont{Klempt}},
  \bibinfo{journal}{Science} \textbf{\bibinfo{volume}{360}},
  \bibinfo{pages}{416} (\bibinfo{year}{2018}).

\bibitem[{\citenamefont{Kunkel et~al.}(2018)\citenamefont{Kunkel, Pr{\"u}fer,
  Strobel, Linnemann, Fr{\"o}lian, Gasenzer, G{\"a}rttner, and
  Oberthaler}}]{kunkel2018spatially}
\bibinfo{author}{\bibfnamefont{P.}~\bibnamefont{Kunkel}},
  \bibinfo{author}{\bibfnamefont{M.}~\bibnamefont{Pr{\"u}fer}},
  \bibinfo{author}{\bibfnamefont{H.}~\bibnamefont{Strobel}},
  \bibinfo{author}{\bibfnamefont{D.}~\bibnamefont{Linnemann}},
  \bibinfo{author}{\bibfnamefont{A.}~\bibnamefont{Fr{\"o}lian}},
  \bibinfo{author}{\bibfnamefont{T.}~\bibnamefont{Gasenzer}},
  \bibinfo{author}{\bibfnamefont{M.}~\bibnamefont{G{\"a}rttner}},
  \bibnamefont{and} \bibinfo{author}{\bibfnamefont{M.~K.}
  \bibnamefont{Oberthaler}}, \bibinfo{journal}{Science}
  \textbf{\bibinfo{volume}{360}}, \bibinfo{pages}{413} (\bibinfo{year}{2018}).

\bibitem[{\citenamefont{Fadel et~al.}(2018)\citenamefont{Fadel, Zibold,
  D{\'e}camps, and Treutlein}}]{fadel2018spatial}
\bibinfo{author}{\bibfnamefont{M.}~\bibnamefont{Fadel}},
  \bibinfo{author}{\bibfnamefont{T.}~\bibnamefont{Zibold}},
  \bibinfo{author}{\bibfnamefont{B.}~\bibnamefont{D{\'e}camps}},
  \bibnamefont{and}
  \bibinfo{author}{\bibfnamefont{P.}~\bibnamefont{Treutlein}},
  \bibinfo{journal}{Science} \textbf{\bibinfo{volume}{360}},
  \bibinfo{pages}{409} (\bibinfo{year}{2018}).

\bibitem[{\citenamefont{Pyrkov and Byrnes}(2013)}]{pyrkov2013}
\bibinfo{author}{\bibfnamefont{A.~N.} \bibnamefont{Pyrkov}} \bibnamefont{and}
  \bibinfo{author}{\bibfnamefont{T.}~\bibnamefont{Byrnes}},
  \bibinfo{journal}{New Journal of Physics} \textbf{\bibinfo{volume}{15}},
  \bibinfo{pages}{093019} (\bibinfo{year}{2013}).

\bibitem[{\citenamefont{Hussain et~al.}(2014)\citenamefont{Hussain, Ilo-Okeke,
  and Byrnes}}]{hussain2014}
\bibinfo{author}{\bibfnamefont{M.~I.} \bibnamefont{Hussain}},
  \bibinfo{author}{\bibfnamefont{E.~O.} \bibnamefont{Ilo-Okeke}},
  \bibnamefont{and} \bibinfo{author}{\bibfnamefont{T.}~\bibnamefont{Byrnes}},
  \bibinfo{journal}{Physical Review A} \textbf{\bibinfo{volume}{89}},
  \bibinfo{pages}{053607} (\bibinfo{year}{2014}).

\bibitem[{\citenamefont{Abdelrahman et~al.}(2014)\citenamefont{Abdelrahman,
  Mukai, H{\"a}ffner, and Byrnes}}]{abdelrahman2014coherent}
\bibinfo{author}{\bibfnamefont{A.}~\bibnamefont{Abdelrahman}},
  \bibinfo{author}{\bibfnamefont{T.}~\bibnamefont{Mukai}},
  \bibinfo{author}{\bibfnamefont{H.}~\bibnamefont{H{\"a}ffner}},
  \bibnamefont{and} \bibinfo{author}{\bibfnamefont{T.}~\bibnamefont{Byrnes}},
  \bibinfo{journal}{Optics express} \textbf{\bibinfo{volume}{22}},
  \bibinfo{pages}{3501} (\bibinfo{year}{2014}).

\bibitem[{\citenamefont{Treutlein et~al.}(2006)\citenamefont{Treutlein,
  H\"ansch, Reichel, Negretti, Cirone, and Calarco}}]{treutlein2006}
\bibinfo{author}{\bibfnamefont{P.}~\bibnamefont{Treutlein}},
  \bibinfo{author}{\bibfnamefont{T.~W.} \bibnamefont{H\"ansch}},
  \bibinfo{author}{\bibfnamefont{J.}~\bibnamefont{Reichel}},
  \bibinfo{author}{\bibfnamefont{A.}~\bibnamefont{Negretti}},
  \bibinfo{author}{\bibfnamefont{M.~A.} \bibnamefont{Cirone}},
  \bibnamefont{and} \bibinfo{author}{\bibfnamefont{T.}~\bibnamefont{Calarco}},
  \bibinfo{journal}{Phys. Rev. A} \textbf{\bibinfo{volume}{74}},
  \bibinfo{pages}{022312} (\bibinfo{year}{2006}).

\bibitem[{\citenamefont{Idlas et~al.}(2016)\citenamefont{Idlas, Domenzain,
  Spreeuw, and Byrnes}}]{idlas2016}
\bibinfo{author}{\bibfnamefont{S.}~\bibnamefont{Idlas}},
  \bibinfo{author}{\bibfnamefont{L.}~\bibnamefont{Domenzain}},
  \bibinfo{author}{\bibfnamefont{R.}~\bibnamefont{Spreeuw}}, \bibnamefont{and}
  \bibinfo{author}{\bibfnamefont{T.}~\bibnamefont{Byrnes}},
  \bibinfo{journal}{Physical Review A} \textbf{\bibinfo{volume}{93}},
  \bibinfo{pages}{022319} (\bibinfo{year}{2016}).

\bibitem[{\citenamefont{Oudot et~al.}(2017)\citenamefont{Oudot, Bancal,
  Schmied, Treutlein, and Sangouard}}]{oudot2017optimal}
\bibinfo{author}{\bibfnamefont{E.}~\bibnamefont{Oudot}},
  \bibinfo{author}{\bibfnamefont{J.-D.} \bibnamefont{Bancal}},
  \bibinfo{author}{\bibfnamefont{R.}~\bibnamefont{Schmied}},
  \bibinfo{author}{\bibfnamefont{P.}~\bibnamefont{Treutlein}},
  \bibnamefont{and}
  \bibinfo{author}{\bibfnamefont{N.}~\bibnamefont{Sangouard}},
  \bibinfo{journal}{Phys. Rev. A} \textbf{\bibinfo{volume}{95}},
  \bibinfo{pages}{052347} (\bibinfo{year}{2017}).

\bibitem[{\citenamefont{Oblak et~al.}(2005)\citenamefont{Oblak, Petrov, Alzar,
  Tittel, Vershovski, Mikkelsen, S{\o}rensen, and Polzik}}]{oblak2005quantum}
\bibinfo{author}{\bibfnamefont{D.}~\bibnamefont{Oblak}},
  \bibinfo{author}{\bibfnamefont{P.~G.} \bibnamefont{Petrov}},
  \bibinfo{author}{\bibfnamefont{C.~L.~G.} \bibnamefont{Alzar}},
  \bibinfo{author}{\bibfnamefont{W.}~\bibnamefont{Tittel}},
  \bibinfo{author}{\bibfnamefont{A.~K.} \bibnamefont{Vershovski}},
  \bibinfo{author}{\bibfnamefont{J.~K.} \bibnamefont{Mikkelsen}},
  \bibinfo{author}{\bibfnamefont{J.~L.} \bibnamefont{S{\o}rensen}},
  \bibnamefont{and} \bibinfo{author}{\bibfnamefont{E.~S.}
  \bibnamefont{Polzik}}, \bibinfo{journal}{Physical Review A}
  \textbf{\bibinfo{volume}{71}}, \bibinfo{pages}{043807}
  (\bibinfo{year}{2005}).

\bibitem[{\citenamefont{Di~Lisi and M{\o}lmer}(2002)}]{di2002entanglement}
\bibinfo{author}{\bibfnamefont{A.}~\bibnamefont{Di~Lisi}} \bibnamefont{and}
  \bibinfo{author}{\bibfnamefont{K.}~\bibnamefont{M{\o}lmer}},
  \bibinfo{journal}{Physical Review A} \textbf{\bibinfo{volume}{66}},
  \bibinfo{pages}{052303} (\bibinfo{year}{2002}).

\bibitem[{\citenamefont{Wang et~al.}(2016)\citenamefont{Wang, Gao, Reinhold,
  Heeres, Ofek, Chou, Axline, Reagor, Blumoff, Sliwa
  et~al.}}]{wang2016schrodinger}
\bibinfo{author}{\bibfnamefont{C.}~\bibnamefont{Wang}},
  \bibinfo{author}{\bibfnamefont{Y.~Y.} \bibnamefont{Gao}},
  \bibinfo{author}{\bibfnamefont{P.}~\bibnamefont{Reinhold}},
  \bibinfo{author}{\bibfnamefont{R.~W.} \bibnamefont{Heeres}},
  \bibinfo{author}{\bibfnamefont{N.}~\bibnamefont{Ofek}},
  \bibinfo{author}{\bibfnamefont{K.}~\bibnamefont{Chou}},
  \bibinfo{author}{\bibfnamefont{C.}~\bibnamefont{Axline}},
  \bibinfo{author}{\bibfnamefont{M.}~\bibnamefont{Reagor}},
  \bibinfo{author}{\bibfnamefont{J.}~\bibnamefont{Blumoff}},
  \bibinfo{author}{\bibfnamefont{K.}~\bibnamefont{Sliwa}},
  \bibnamefont{et~al.}, \bibinfo{journal}{Science}
  \textbf{\bibinfo{volume}{352}}, \bibinfo{pages}{1087} (\bibinfo{year}{2016}).

\bibitem[{\citenamefont{Julsgaard
  et~al.}(2001{\natexlab{b}})\citenamefont{Julsgaard, Kozhekin, and
  Polzik}}]{julsgaard2001experimental}
\bibinfo{author}{\bibfnamefont{B.}~\bibnamefont{Julsgaard}},
  \bibinfo{author}{\bibfnamefont{A.}~\bibnamefont{Kozhekin}}, \bibnamefont{and}
  \bibinfo{author}{\bibfnamefont{E.~S.} \bibnamefont{Polzik}},
  \bibinfo{journal}{Nature} \textbf{\bibinfo{volume}{413}},
  \bibinfo{pages}{400} (\bibinfo{year}{2001}{\natexlab{b}}).

\bibitem[{\citenamefont{Kuzmich et~al.}(2000)\citenamefont{Kuzmich, Mandel, and
  Bigelow}}]{kuzmich2000generation}
\bibinfo{author}{\bibfnamefont{A.}~\bibnamefont{Kuzmich}},
  \bibinfo{author}{\bibfnamefont{L.}~\bibnamefont{Mandel}}, \bibnamefont{and}
  \bibinfo{author}{\bibfnamefont{N.}~\bibnamefont{Bigelow}},
  \bibinfo{journal}{Physical Review Letters} \textbf{\bibinfo{volume}{85}},
  \bibinfo{pages}{1594} (\bibinfo{year}{2000}).

\bibitem[{\citenamefont{Chou et~al.}(2005)\citenamefont{Chou, De~Riedmatten,
  Felinto, Polyakov, Van~Enk, and Kimble}}]{chou2005measurement}
\bibinfo{author}{\bibfnamefont{C.-W.} \bibnamefont{Chou}},
  \bibinfo{author}{\bibfnamefont{H.}~\bibnamefont{De~Riedmatten}},
  \bibinfo{author}{\bibfnamefont{D.}~\bibnamefont{Felinto}},
  \bibinfo{author}{\bibfnamefont{S.~V.} \bibnamefont{Polyakov}},
  \bibinfo{author}{\bibfnamefont{S.~J.} \bibnamefont{Van~Enk}},
  \bibnamefont{and} \bibinfo{author}{\bibfnamefont{H.~J.}
  \bibnamefont{Kimble}}, \bibinfo{journal}{Nature}
  \textbf{\bibinfo{volume}{438}}, \bibinfo{pages}{828} (\bibinfo{year}{2005}).

\bibitem[{\citenamefont{Muschik et~al.}(2011)\citenamefont{Muschik, Polzik, and
  Cirac}}]{muschik2011dissipatively}
\bibinfo{author}{\bibfnamefont{C.~A.} \bibnamefont{Muschik}},
  \bibinfo{author}{\bibfnamefont{E.~S.} \bibnamefont{Polzik}},
  \bibnamefont{and} \bibinfo{author}{\bibfnamefont{J.~I.} \bibnamefont{Cirac}},
  \bibinfo{journal}{Physical Review A} \textbf{\bibinfo{volume}{83}},
  \bibinfo{pages}{052312} (\bibinfo{year}{2011}).

\bibitem[{\citenamefont{Duan et~al.}(2000{\natexlab{a}})\citenamefont{Duan,
  Cirac, Zoller, and Polzik}}]{PhysRevLett.85.5643}
\bibinfo{author}{\bibfnamefont{L.-M.} \bibnamefont{Duan}},
  \bibinfo{author}{\bibfnamefont{J.~I.} \bibnamefont{Cirac}},
  \bibinfo{author}{\bibfnamefont{P.}~\bibnamefont{Zoller}}, \bibnamefont{and}
  \bibinfo{author}{\bibfnamefont{E.~S.} \bibnamefont{Polzik}},
  \bibinfo{journal}{Phys. Rev. Lett.} \textbf{\bibinfo{volume}{85}},
  \bibinfo{pages}{5643} (\bibinfo{year}{2000}{\natexlab{a}}).

\bibitem[{\citenamefont{Pettersson and Byrnes}(2017)}]{pettersson2017light}
\bibinfo{author}{\bibfnamefont{O.}~\bibnamefont{Pettersson}} \bibnamefont{and}
  \bibinfo{author}{\bibfnamefont{T.}~\bibnamefont{Byrnes}},
  \bibinfo{journal}{Physical Review A} \textbf{\bibinfo{volume}{95}},
  \bibinfo{pages}{043817} (\bibinfo{year}{2017}).

\bibitem[{\citenamefont{Asitzabal-Zuluana
  et~al.}(2020)\citenamefont{Asitzabal-Zuluana, Skobleva, Richter, Ji, Mao,
  Kondappan, Ivannikov, and Byrnes}}]{juanQND}
\bibinfo{author}{\bibfnamefont{J.}~\bibnamefont{Asitzabal-Zuluana}},
  \bibinfo{author}{\bibfnamefont{I.}~\bibnamefont{Skobleva}},
  \bibinfo{author}{\bibfnamefont{L.}~\bibnamefont{Richter}},
  \bibinfo{author}{\bibfnamefont{Y.}~\bibnamefont{Ji}},
  \bibinfo{author}{\bibfnamefont{Y.}~\bibnamefont{Mao}},
  \bibinfo{author}{\bibfnamefont{M.}~\bibnamefont{Kondappan}},
  \bibinfo{author}{\bibfnamefont{V.}~\bibnamefont{Ivannikov}},
  \bibnamefont{and} \bibinfo{author}{\bibfnamefont{T.}~\bibnamefont{Byrnes}},
  \textbf{\bibinfo{volume}{409}} (\bibinfo{year}{2020}).

\bibitem[{\citenamefont{Pyrkov and Byrnes}(2014)}]{pyrkov2014quantum}
\bibinfo{author}{\bibfnamefont{A.~N.} \bibnamefont{Pyrkov}} \bibnamefont{and}
  \bibinfo{author}{\bibfnamefont{T.}~\bibnamefont{Byrnes}},
  \bibinfo{journal}{New Journal of Physics} \textbf{\bibinfo{volume}{16}},
  \bibinfo{pages}{073038} (\bibinfo{year}{2014}).

\bibitem[{\citenamefont{Ilo-Okeke et~al.}(2018)\citenamefont{Ilo-Okeke,
  Tessler, Dowling, and Byrnes}}]{ilo2018remote}
\bibinfo{author}{\bibfnamefont{E.~O.} \bibnamefont{Ilo-Okeke}},
  \bibinfo{author}{\bibfnamefont{L.}~\bibnamefont{Tessler}},
  \bibinfo{author}{\bibfnamefont{J.~P.} \bibnamefont{Dowling}},
  \bibnamefont{and} \bibinfo{author}{\bibfnamefont{T.}~\bibnamefont{Byrnes}},
  \bibinfo{journal}{npj Quantum Information} \textbf{\bibinfo{volume}{4}},
  \bibinfo{pages}{1} (\bibinfo{year}{2018}).

\bibitem[{\citenamefont{Chaudhary et~al.}(2021)\citenamefont{Chaudhary, Fadel,
  Ilo-Okeke, Pyrkov, Ivannikov, and Byrnes}}]{manish2021}
\bibinfo{author}{\bibfnamefont{M.}~\bibnamefont{Chaudhary}},
  \bibinfo{author}{\bibfnamefont{M.}~\bibnamefont{Fadel}},
  \bibinfo{author}{\bibfnamefont{E.~O.} \bibnamefont{Ilo-Okeke}},
  \bibinfo{author}{\bibfnamefont{A.~N.} \bibnamefont{Pyrkov}},
  \bibinfo{author}{\bibfnamefont{V.}~\bibnamefont{Ivannikov}},
  \bibnamefont{and} \bibinfo{author}{\bibfnamefont{T.}~\bibnamefont{Byrnes}},
  \bibinfo{journal}{Phys. Rev. A} \textbf{\bibinfo{volume}{103}},
  \bibinfo{pages}{062417} (\bibinfo{year}{2021}).

\bibitem[{\citenamefont{Windpassinger et~al.}(2008)\citenamefont{Windpassinger,
  Oblak, Petrov, Kubasik, Saffman, Alzar, Appel, M{\"u}ller, Kj{\ae}rgaard, and
  Polzik}}]{windpassinger2008nondestructive}
\bibinfo{author}{\bibfnamefont{P.}~\bibnamefont{Windpassinger}},
  \bibinfo{author}{\bibfnamefont{D.}~\bibnamefont{Oblak}},
  \bibinfo{author}{\bibfnamefont{P.}~\bibnamefont{Petrov}},
  \bibinfo{author}{\bibfnamefont{M.}~\bibnamefont{Kubasik}},
  \bibinfo{author}{\bibfnamefont{M.}~\bibnamefont{Saffman}},
  \bibinfo{author}{\bibfnamefont{C.~G.} \bibnamefont{Alzar}},
  \bibinfo{author}{\bibfnamefont{J.}~\bibnamefont{Appel}},
  \bibinfo{author}{\bibfnamefont{J.}~\bibnamefont{M{\"u}ller}},
  \bibinfo{author}{\bibfnamefont{N.}~\bibnamefont{Kj{\ae}rgaard}},
  \bibnamefont{and} \bibinfo{author}{\bibfnamefont{E.}~\bibnamefont{Polzik}},
  \bibinfo{journal}{Physical review letters} \textbf{\bibinfo{volume}{100}},
  \bibinfo{pages}{103601} (\bibinfo{year}{2008}).

\bibitem[{\citenamefont{Meiser et~al.}(2008)\citenamefont{Meiser, Ye, and
  Holland}}]{meiser2008spin}
\bibinfo{author}{\bibfnamefont{D.}~\bibnamefont{Meiser}},
  \bibinfo{author}{\bibfnamefont{J.}~\bibnamefont{Ye}}, \bibnamefont{and}
  \bibinfo{author}{\bibfnamefont{M.}~\bibnamefont{Holland}},
  \bibinfo{journal}{New Journal of Physics} \textbf{\bibinfo{volume}{10}},
  \bibinfo{pages}{073014} (\bibinfo{year}{2008}).

\bibitem[{\citenamefont{Byrnes et~al.}(2012)\citenamefont{Byrnes, Wen, and
  Yamamoto}}]{byrnes2012macroscopic}
\bibinfo{author}{\bibfnamefont{T.}~\bibnamefont{Byrnes}},
  \bibinfo{author}{\bibfnamefont{K.}~\bibnamefont{Wen}}, \bibnamefont{and}
  \bibinfo{author}{\bibfnamefont{Y.}~\bibnamefont{Yamamoto}},
  \bibinfo{journal}{Physical Review A} \textbf{\bibinfo{volume}{85}},
  \bibinfo{pages}{040306} (\bibinfo{year}{2012}).

\bibitem[{\citenamefont{Reichel and Vuletic}(2011)}]{reichel2011atom}
\bibinfo{author}{\bibfnamefont{J.}~\bibnamefont{Reichel}} \bibnamefont{and}
  \bibinfo{author}{\bibfnamefont{V.}~\bibnamefont{Vuletic}},
  \emph{\bibinfo{title}{Atom chips}} (\bibinfo{publisher}{John Wiley \& Sons},
  \bibinfo{year}{2011}).

\bibitem[{\citenamefont{Whitlock and Gerritsma}(2009)}]{whitlock2009t}
\bibinfo{author}{\bibfnamefont{S.}~\bibnamefont{Whitlock}} \bibnamefont{and}
  \bibinfo{author}{\bibfnamefont{R.}~\bibnamefont{Gerritsma}},
  \bibinfo{journal}{New Journal of Physics} \textbf{\bibinfo{volume}{11}},
  \bibinfo{pages}{023021} (\bibinfo{year}{2009}).

\bibitem[{\citenamefont{Geremia et~al.}(2006)\citenamefont{Geremia, Stockton,
  and Mabuchi}}]{PhysRevA.73.042112}
\bibinfo{author}{\bibfnamefont{J.~M.} \bibnamefont{Geremia}},
  \bibinfo{author}{\bibfnamefont{J.~K.} \bibnamefont{Stockton}},
  \bibnamefont{and} \bibinfo{author}{\bibfnamefont{H.}~\bibnamefont{Mabuchi}},
  \bibinfo{journal}{Phys. Rev. A} \textbf{\bibinfo{volume}{73}},
  \bibinfo{pages}{042112} (\bibinfo{year}{2006}).

\bibitem[{\citenamefont{Kubasik et~al.}(2009)\citenamefont{Kubasik,
  Koschorreck, Napolitano, de~Echaniz, Crepaz, Eschner, Polzik, and
  Mitchell}}]{PhysRevA.79.043815}
\bibinfo{author}{\bibfnamefont{M.}~\bibnamefont{Kubasik}},
  \bibinfo{author}{\bibfnamefont{M.}~\bibnamefont{Koschorreck}},
  \bibinfo{author}{\bibfnamefont{M.}~\bibnamefont{Napolitano}},
  \bibinfo{author}{\bibfnamefont{S.~R.} \bibnamefont{de~Echaniz}},
  \bibinfo{author}{\bibfnamefont{H.}~\bibnamefont{Crepaz}},
  \bibinfo{author}{\bibfnamefont{J.}~\bibnamefont{Eschner}},
  \bibinfo{author}{\bibfnamefont{E.~S.} \bibnamefont{Polzik}},
  \bibnamefont{and} \bibinfo{author}{\bibfnamefont{M.~W.}
  \bibnamefont{Mitchell}}, \bibinfo{journal}{Phys. Rev. A}
  \textbf{\bibinfo{volume}{79}}, \bibinfo{pages}{043815}
  (\bibinfo{year}{2009}).

\bibitem[{\citenamefont{de~Echaniz et~al.}(2005)\citenamefont{de~Echaniz,
  Mitchell, Kubasik, Koschorreck, Crepaz, Eschner, and Polzik}}]{Echaniz_2005}
\bibinfo{author}{\bibfnamefont{S.~R.} \bibnamefont{de~Echaniz}},
  \bibinfo{author}{\bibfnamefont{M.~W.} \bibnamefont{Mitchell}},
  \bibinfo{author}{\bibfnamefont{M.}~\bibnamefont{Kubasik}},
  \bibinfo{author}{\bibfnamefont{M.}~\bibnamefont{Koschorreck}},
  \bibinfo{author}{\bibfnamefont{H.}~\bibnamefont{Crepaz}},
  \bibinfo{author}{\bibfnamefont{J.}~\bibnamefont{Eschner}}, \bibnamefont{and}
  \bibinfo{author}{\bibfnamefont{E.~S.} \bibnamefont{Polzik}},
  \bibinfo{journal}{Journal of Optics B: Quantum and Semiclassical Optics}
  \textbf{\bibinfo{volume}{7}}, \bibinfo{pages}{S548} (\bibinfo{year}{2005}).

\bibitem[{\citenamefont{Ilo-Okeke and Byrnes}(2014)}]{ilo2014theory}
\bibinfo{author}{\bibfnamefont{E.~O.} \bibnamefont{Ilo-Okeke}}
  \bibnamefont{and} \bibinfo{author}{\bibfnamefont{T.}~\bibnamefont{Byrnes}},
  \bibinfo{journal}{Physical review letters} \textbf{\bibinfo{volume}{112}},
  \bibinfo{pages}{233602} (\bibinfo{year}{2014}).

\bibitem[{\citenamefont{Ilo-Okeke and Byrnes}(2016)}]{PhysRevA.94.013617}
\bibinfo{author}{\bibfnamefont{E.~O.} \bibnamefont{Ilo-Okeke}}
  \bibnamefont{and} \bibinfo{author}{\bibfnamefont{T.}~\bibnamefont{Byrnes}},
  \bibinfo{journal}{Phys. Rev. A} \textbf{\bibinfo{volume}{94}},
  \bibinfo{pages}{013617} (\bibinfo{year}{2016}).

\bibitem[{\citenamefont{Duan et~al.}(2000{\natexlab{b}})\citenamefont{Duan,
  Cirac, Zoller, and Polzik}}]{duan2000quantum}
\bibinfo{author}{\bibfnamefont{L.-M.} \bibnamefont{Duan}},
  \bibinfo{author}{\bibfnamefont{J.}~\bibnamefont{Cirac}},
  \bibinfo{author}{\bibfnamefont{P.}~\bibnamefont{Zoller}}, \bibnamefont{and}
  \bibinfo{author}{\bibfnamefont{E.}~\bibnamefont{Polzik}},
  \bibinfo{journal}{Physical review letters} \textbf{\bibinfo{volume}{85}},
  \bibinfo{pages}{5643} (\bibinfo{year}{2000}{\natexlab{b}}).

\bibitem[{\citenamefont{Serafin et~al.}(2021)\citenamefont{Serafin, Fadel,
  Treutlein, and Sinatra}}]{serafin2021nuclear}
\bibinfo{author}{\bibfnamefont{A.}~\bibnamefont{Serafin}},
  \bibinfo{author}{\bibfnamefont{M.}~\bibnamefont{Fadel}},
  \bibinfo{author}{\bibfnamefont{P.}~\bibnamefont{Treutlein}},
  \bibnamefont{and} \bibinfo{author}{\bibfnamefont{A.}~\bibnamefont{Sinatra}},
  \bibinfo{journal}{Physical Review Letters} \textbf{\bibinfo{volume}{127}},
  \bibinfo{pages}{013601} (\bibinfo{year}{2021}).

\bibitem[{\citenamefont{Tsang and Caves}(2012)}]{tsang2012evading}
\bibinfo{author}{\bibfnamefont{M.}~\bibnamefont{Tsang}} \bibnamefont{and}
  \bibinfo{author}{\bibfnamefont{C.~M.} \bibnamefont{Caves}},
  \bibinfo{journal}{Physical Review X} \textbf{\bibinfo{volume}{2}},
  \bibinfo{pages}{031016} (\bibinfo{year}{2012}).

\bibitem[{\citenamefont{Lone and Byrnes}(2015)}]{acstark}
\bibinfo{author}{\bibfnamefont{M.~Q.} \bibnamefont{Lone}} \bibnamefont{and}
  \bibinfo{author}{\bibfnamefont{T.}~\bibnamefont{Byrnes}},
  \bibinfo{journal}{Phys. Rev. A} \textbf{\bibinfo{volume}{92}},
  \bibinfo{pages}{011401} (\bibinfo{year}{2015}).

\bibitem[{\citenamefont{Autler and Townes}(1955)}]{autler1955stark}
\bibinfo{author}{\bibfnamefont{S.~H.} \bibnamefont{Autler}} \bibnamefont{and}
  \bibinfo{author}{\bibfnamefont{C.~H.} \bibnamefont{Townes}},
  \bibinfo{journal}{Physical Review} \textbf{\bibinfo{volume}{100}},
  \bibinfo{pages}{703} (\bibinfo{year}{1955}).

\bibitem[{\citenamefont{Gehm et~al.}(1998)\citenamefont{Gehm, O'Hara, Savard,
  and Thomas}}]{atomtrapnoise}
\bibinfo{author}{\bibfnamefont{M.~E.} \bibnamefont{Gehm}},
  \bibinfo{author}{\bibfnamefont{K.~M.} \bibnamefont{O'Hara}},
  \bibinfo{author}{\bibfnamefont{T.~A.} \bibnamefont{Savard}},
  \bibnamefont{and} \bibinfo{author}{\bibfnamefont{J.~E.}
  \bibnamefont{Thomas}}, \bibinfo{journal}{Phys. Rev. A}
  \textbf{\bibinfo{volume}{58}}, \bibinfo{pages}{3914} (\bibinfo{year}{1998}).

\bibitem[{\citenamefont{Byrnes et~al.}(2015)\citenamefont{Byrnes, Rosseau,
  Khosla, Pyrkov, Thomasen, Mukai, Koyama, Abdelrahman, and
  Ilo-Okeke}}]{byrnes2015macroscopic}
\bibinfo{author}{\bibfnamefont{T.}~\bibnamefont{Byrnes}},
  \bibinfo{author}{\bibfnamefont{D.}~\bibnamefont{Rosseau}},
  \bibinfo{author}{\bibfnamefont{M.}~\bibnamefont{Khosla}},
  \bibinfo{author}{\bibfnamefont{A.}~\bibnamefont{Pyrkov}},
  \bibinfo{author}{\bibfnamefont{A.}~\bibnamefont{Thomasen}},
  \bibinfo{author}{\bibfnamefont{T.}~\bibnamefont{Mukai}},
  \bibinfo{author}{\bibfnamefont{S.}~\bibnamefont{Koyama}},
  \bibinfo{author}{\bibfnamefont{A.}~\bibnamefont{Abdelrahman}},
  \bibnamefont{and}
  \bibinfo{author}{\bibfnamefont{E.}~\bibnamefont{Ilo-Okeke}},
  \bibinfo{journal}{Optics Communications} \textbf{\bibinfo{volume}{337}},
  \bibinfo{pages}{102} (\bibinfo{year}{2015}).

\bibitem[{\citenamefont{Vidal and Werner}(2002)}]{vidal2002computable}
\bibinfo{author}{\bibfnamefont{G.}~\bibnamefont{Vidal}} \bibnamefont{and}
  \bibinfo{author}{\bibfnamefont{R.~F.} \bibnamefont{Werner}},
  \bibinfo{journal}{Physical Review A} \textbf{\bibinfo{volume}{65}},
  \bibinfo{pages}{032314} (\bibinfo{year}{2002}).

\bibitem[{\citenamefont{Plenio}(2005)}]{plenio2005logarithmic}
\bibinfo{author}{\bibfnamefont{M.~B.} \bibnamefont{Plenio}},
  \bibinfo{journal}{Physical review letters} \textbf{\bibinfo{volume}{95}},
  \bibinfo{pages}{090503} (\bibinfo{year}{2005}).

\bibitem[{\citenamefont{Hofmann and Takeuchi}(2003)}]{hofmann2003violation}
\bibinfo{author}{\bibfnamefont{H.~F.} \bibnamefont{Hofmann}} \bibnamefont{and}
  \bibinfo{author}{\bibfnamefont{S.}~\bibnamefont{Takeuchi}},
  \bibinfo{journal}{Physical Review A} \textbf{\bibinfo{volume}{68}},
  \bibinfo{pages}{032103} (\bibinfo{year}{2003}).

\bibitem[{\citenamefont{Sun et~al.}(2018)\citenamefont{Sun, Ye, Xiao, Xu, Wu,
  Xu, Chen, Li, and Guo}}]{Sun2018}
\bibinfo{author}{\bibfnamefont{K.}~\bibnamefont{Sun}},
  \bibinfo{author}{\bibfnamefont{X.-J.} \bibnamefont{Ye}},
  \bibinfo{author}{\bibfnamefont{Y.}~\bibnamefont{Xiao}},
  \bibinfo{author}{\bibfnamefont{X.-Y.} \bibnamefont{Xu}},
  \bibinfo{author}{\bibfnamefont{Y.-C.} \bibnamefont{Wu}},
  \bibinfo{author}{\bibfnamefont{J.-S.} \bibnamefont{Xu}},
  \bibinfo{author}{\bibfnamefont{J.-L.} \bibnamefont{Chen}},
  \bibinfo{author}{\bibfnamefont{C.-F.} \bibnamefont{Li}}, \bibnamefont{and}
  \bibinfo{author}{\bibfnamefont{G.-C.} \bibnamefont{Guo}},
  \bibinfo{journal}{npj Quantum Information} \textbf{\bibinfo{volume}{4}},
  \bibinfo{pages}{12} (\bibinfo{year}{2018}), ISSN \bibinfo{issn}{2056-6387}.

\bibitem[{\citenamefont{Reid et~al.}(2009)\citenamefont{Reid, Drummond, Bowen,
  Cavalcanti, Lam, Bachor, Andersen, and Leuchs}}]{reid2009colloquium}
\bibinfo{author}{\bibfnamefont{M.}~\bibnamefont{Reid}},
  \bibinfo{author}{\bibfnamefont{P.}~\bibnamefont{Drummond}},
  \bibinfo{author}{\bibfnamefont{W.}~\bibnamefont{Bowen}},
  \bibinfo{author}{\bibfnamefont{E.~G.} \bibnamefont{Cavalcanti}},
  \bibinfo{author}{\bibfnamefont{P.~K.} \bibnamefont{Lam}},
  \bibinfo{author}{\bibfnamefont{H.}~\bibnamefont{Bachor}},
  \bibinfo{author}{\bibfnamefont{U.~L.} \bibnamefont{Andersen}},
  \bibnamefont{and} \bibinfo{author}{\bibfnamefont{G.}~\bibnamefont{Leuchs}},
  \bibinfo{journal}{Reviews of Modern Physics} \textbf{\bibinfo{volume}{81}},
  \bibinfo{pages}{1727} (\bibinfo{year}{2009}).

\bibitem[{\citenamefont{Adesso et~al.}(2016)\citenamefont{Adesso, Bromley, and
  Cianciaruso}}]{adesso2016measures}
\bibinfo{author}{\bibfnamefont{G.}~\bibnamefont{Adesso}},
  \bibinfo{author}{\bibfnamefont{T.~R.} \bibnamefont{Bromley}},
  \bibnamefont{and}
  \bibinfo{author}{\bibfnamefont{M.}~\bibnamefont{Cianciaruso}},
  \bibinfo{journal}{Journal of Physics A: Mathematical and Theoretical}
  \textbf{\bibinfo{volume}{49}}, \bibinfo{pages}{473001}
  (\bibinfo{year}{2016}).

\bibitem[{\citenamefont{Ma et~al.}(2019)\citenamefont{Ma, Cui, Cao, Fei,
  Vedral, Byrnes, and Radhakrishnan}}]{ma2019operational}
\bibinfo{author}{\bibfnamefont{Z.-H.} \bibnamefont{Ma}},
  \bibinfo{author}{\bibfnamefont{J.}~\bibnamefont{Cui}},
  \bibinfo{author}{\bibfnamefont{Z.}~\bibnamefont{Cao}},
  \bibinfo{author}{\bibfnamefont{S.-M.} \bibnamefont{Fei}},
  \bibinfo{author}{\bibfnamefont{V.}~\bibnamefont{Vedral}},
  \bibinfo{author}{\bibfnamefont{T.}~\bibnamefont{Byrnes}}, \bibnamefont{and}
  \bibinfo{author}{\bibfnamefont{C.}~\bibnamefont{Radhakrishnan}},
  \bibinfo{journal}{EPL (Europhysics Letters)} \textbf{\bibinfo{volume}{125}},
  \bibinfo{pages}{50005} (\bibinfo{year}{2019}).

\bibitem[{\citenamefont{B{\"o}hi et~al.}(2009)\citenamefont{B{\"o}hi, Riedel,
  Hoffrogge, Reichel, H{\"a}nsch, and Treutlein}}]{bohi2009coherent}
\bibinfo{author}{\bibfnamefont{P.}~\bibnamefont{B{\"o}hi}},
  \bibinfo{author}{\bibfnamefont{M.~F.} \bibnamefont{Riedel}},
  \bibinfo{author}{\bibfnamefont{J.}~\bibnamefont{Hoffrogge}},
  \bibinfo{author}{\bibfnamefont{J.}~\bibnamefont{Reichel}},
  \bibinfo{author}{\bibfnamefont{T.~W.} \bibnamefont{H{\"a}nsch}},
  \bibnamefont{and}
  \bibinfo{author}{\bibfnamefont{P.}~\bibnamefont{Treutlein}},
  \bibinfo{journal}{Nature Physics} \textbf{\bibinfo{volume}{5}},
  \bibinfo{pages}{592} (\bibinfo{year}{2009}).

\bibitem[{\citenamefont{Schleier-Smith
  et~al.}(2010)\citenamefont{Schleier-Smith, Leroux, and
  Vuleti{\'c}}}]{schleier2010states}
\bibinfo{author}{\bibfnamefont{M.~H.} \bibnamefont{Schleier-Smith}},
  \bibinfo{author}{\bibfnamefont{I.~D.} \bibnamefont{Leroux}},
  \bibnamefont{and}
  \bibinfo{author}{\bibfnamefont{V.}~\bibnamefont{Vuleti{\'c}}},
  \bibinfo{journal}{Physical review letters} \textbf{\bibinfo{volume}{104}},
  \bibinfo{pages}{073604} (\bibinfo{year}{2010}).

\bibitem[{\citenamefont{Julienne et~al.}(1997)\citenamefont{Julienne, Mies,
  Tiesinga, and Williams}}]{julienne1997}
\bibinfo{author}{\bibfnamefont{P.~S.} \bibnamefont{Julienne}},
  \bibinfo{author}{\bibfnamefont{F.~H.} \bibnamefont{Mies}},
  \bibinfo{author}{\bibfnamefont{E.}~\bibnamefont{Tiesinga}}, \bibnamefont{and}
  \bibinfo{author}{\bibfnamefont{C.~J.} \bibnamefont{Williams}},
  \bibinfo{journal}{Phys. Rev. Lett.} \textbf{\bibinfo{volume}{78}},
  \bibinfo{pages}{1880} (\bibinfo{year}{1997}).

\bibitem[{\citenamefont{Steck}(2021)}]{steck2021}
\bibinfo{author}{\bibfnamefont{D.~A.} \bibnamefont{Steck}},
  \emph{\bibinfo{title}{{Rubidium 87 D line Data}}} (\bibinfo{year}{2021}),
  \urlprefix\url{http://www.steck.us/alkalidata}.

\end{thebibliography}

\end{document}